\documentclass[a4paper,11pt]{article}
\usepackage{jheppub}

\usepackage{graphicx,dcolumn,bm,amsfonts,amsmath,color,xcolor,amsthm}
\usepackage[colorlinks=true]{hyperref}
\usepackage{xcolor}
\usepackage{natbib}
\usepackage{multirow}
\usepackage{ulem}

\allowdisplaybreaks

\title{Addressing the $S$-wave scalar $f_0(1500)$-resonance in quasi-four-body FCNC rare $B_s \to f_0(1500) (\to \pi^ + \pi^- ) \ell^ + \ell^- / \nu \bar \nu$ decays}
\author{Xue Zheng,}
\author{Hai-Bing Fu$^*$,}
\author{Dan-Dan Hu,}
\author{Jing-Kai Yang,}
\author{Jian-Qi Chen,}
\author{Wan-Bing Luo}

\affiliation{Department of Physics, Guizhou Minzu University, Guiyang 550025, P. R. China}
\emailAdd{fuhb@gzmu.edu.cn}

\abstract{
The $f_0(1500)$ is a quite special resonance in the light scalar  spectrum. Compared with other nearby resonance, it has an unexpectedly narrow decay width. Since its mass region overlaps strongly with $f_0(1370)$ and $f_0(980)$, physicists have long debated its internal structure. To explore this issue, this work adopted the conventional quark-antiquark $q\bar{q}$ picture and investigated the behavior of the $f_0(1500)$-resonance in quasi-four-body decay channels. Based on this, we constructed a twist-2 light-cone distribution amplitude(LCDA) schemes based on the light-cone harmonic oscillator model, and presented their moments $\langle \xi ^n _{2;f_0(1500)} \rangle |_\mu$ and Gegenbauer moments $a_{n;f_0(1500)}(\mu)$ at $\mu_0=1~\mathrm{GeV}$ and $\mu_k= 3~\mathrm{GeV}$ for $n=1,3,5$. Meanwhile, the $B_s\to f_0(1500)$ transition form factors (TFFs) are calculated by using the QCD light-cone sum rule. Then, we obtained the three TFFs at large recoil point, {\it i.e.,} $f_ + ^{B_s f_0(1500)}(0)= 0.390_{-0.046}^{ + 0.047}$, $f_-^{B_s f_0(1500)}(0)= -0.460_{-0.055}^{ + 0.051}$, and $f_{\rm T}^{B_s f_0(1500)}(0)= 0.568^{ + 0.069}_{-0.065}$. In addition, we extrapolated TFFs to the whole physical $q^2$-region by using the simplified $z(q^2)$-series expansion. Then we computed the branching fractions of the quasi-four-body rare decays $B_s \to f_0(1500)(\to \pi^ + \pi^-)\ell^ + \ell^-$ and $B_s \to f_0(1500)(\to \pi^ + \pi^-)\nu\bar{\nu}$. For comparison, we also present the results of the corresponding three-body decays obtained under the narrow-width approximation. Finally, we show the distribution of the double-differential decay width $d^2\Gamma/ds dq^2$ for the quasi-four-body decay $B_s\to f_0(1500)(\to\pi^ + \pi^-)\mu^ + \mu^-$. We hope that our predictions can provide a useful theoretical reference for future experimental measurements and phenomenological research.
}


\begin{document}

\maketitle

\bibliographystyle{JHEP}

\section{Introduction}\label {Sec:I}
In the research of non-perturbative quantum chromodynamics (QCD), figuring out the internal structure and dynamical nature of light scalar resonances remains one of the central open problems. Light scalar states contribute to hadron mass generation via the spontaneous chiral symmetry breaking of QCD and occupy a central position in the evolution of low-energy hadron dynamics. Revealing their quark-gluon substructure is therefore a key step toward understanding color confinement and improving the theoretical framework of non-perturbative QCD~\cite{Pelaez:2015qba}. The $f_0(1500)$, with a mass around 1500 MeV, is a representative light scalar resonance whose composition and physical properties have long been heavily debated. Up to now, existing explanations regarding the internal structure of scalar states include conventional quark-antiquark $q \bar{q}$ states~\cite{Cheng:2005nb}, hybrid states~\cite{Klempt:2021nuf}, molecule states~\cite{Weinstein:1983gd,Weinstein:1982gc}, as well as tetra-quark states~\cite{Jaffe:1976ih} and the superpositions of these contents~\cite{Amsler:1995td, Amsler:1995tu, Amsler:2002ey}. There is no general agreement on the nature of these states~\cite{ParticleDataGroup:2006fqo}, making $f_0(1500)$ a key subject in hadron-physics studies.

At present, both experimental and theoretical groups have been widely exploring the nature of light scalar states by means of high-precision many-body semileptonic decays. In the charmed meson semileptonic decays, the BESIII collaboration has systematically investigated several decay processes, including $D^ + \to f_{0}(500)(\to\pi^ + \pi^{-})e^ + \nu_{e}$, $D^ + \to f_{0}(980)(\to\pi^ + \pi^{-})e^ + \nu_{e}$ and $D^{0}\to a_{0}(980)(\to\eta\pi^{-})e^ + \nu_{e}$~\cite{BESIII:2024lnh,BESIII:2018qmf,BESIII:2024zvp}. These results not only confirmed the existence of scalar resonances, but also provided important experimental evidence for understanding the internal structure of light scalar states. In the sector of bottom-flavored states, theoretical studies on four-body semileptonic decays have also become mature. Analyses based on the SU$_f(3)$ flavor symmetry can effectively relate the amplitudes of different decay channels and yield definite relations between branching fractions~\cite{Wan:2024ssl}. The QCD light-cone sum rules offer a powerful approach to compute TFFs starting from QCD correlation functions together with non-perturbative parameterizations. Furthermore, the Flatt\'e formula, when adopted to describe the propagator of wide resonances, allows a more accurate treatment of the line-shape effects for broad resonances such as $K_{0}^{*}(700)$~\cite{Huang:2026zxp}. All these works demonstrate that the theoretical framework for four-body semileptonic decays is well-established and can provide reliable theoretical guidance for experimental measurements. At present, systematic investigations on quasi-four-body rare decay processes involving $f_0(1500)$ are still rather scarc, with obvious deficiencies in the relevant theoretical calculations and phenomenological analyses that need to be further improved. This provides the main motivation for our present work. We aim to provide definite theoretical predictions and reference information for future experimental explorations.

For scalar quark-antiquark states above $1\,\mathrm{GeV}$, a series of scalar resonances including $f_0(1500)$, $f_0(1370)$, $a_0(1450)$ and $K_0^*(1430)$ have been experimentally observed. Their quantum numbers, dominant decay modes, branching fractions, masses and widths are summarized in Table ~\ref{table:meson information}. Among these states, the $f_0(1500)$-resonance with $J^{PC}=0^{ + + }$ possesses mass $1522\pm25\ \mathrm{MeV}$ and width $108\pm33\ \mathrm{MeV}$, and it mainly decays into $\pi\pi$ and $K\bar{K}$ channels with branching fractions $(34.5\pm2.2)\%$ and $(8.5\pm1.0)\%$~\cite{ParticleDataGroup:2026aaa}, respectively. The broad $f_0(1370)$ has a large width in the range of $200$ to $500\ \mathrm{MeV}$, and its mass parameter is poorly constrained within $1200-1500\ \mathrm{MeV}$. The $a_0(1450)$ and $K_0^*(1430)$ also possess considerable decay widths, which can induce nontrivial channel-channel interference when multi-body final states are considered~\cite{Klempt:2007cp}. Because these scalar resonances above $1\,\mathrm{GeV}$ have overlapping mass ranges and many open decay channels, it is difficult to separate their intrinsic properties and obtain reliable physical parameters from experimental data~\cite{Anisovich:2001zp}. Although the mass and width parameters of these scalar resonances are experimentally known, finite-width effects are frequently underestimated in phenomenological studies of $B_s$-meson quasi-four-body rare decays. For broad resonances like $f_0(1370)$ and $f_0(1500)$, their widths are much larger than normal narrow states. For these broad states, the usual simple methods like the narrow-width approximation, the $\delta(s-M^2)$ description, and the constant-width Breit-Wigner propagator are no longer reliable. They can cause large theoretical errors.
Also, $f_0(1370)$ and $f_0(1500)$ have overlapping masses. Their resonance tails will interfere with each other in the $\pi\pi$ and $K\bar{K}$ invariant mass spectra. Meanwhile, $a_0(1450)$ and $K_0^{*}(1430)$ will add extra coherent backgrounds. In this multi-resonance situation, finite-width effects are more than just simple peak broadening they amplify coupled-channel interference. If we treat each resonance separately under the narrow-width approximation, we will lose important phase information and lower the credibility of our theoretical decay predictions.
\begin{table}[htb]
\footnotesize
\begin{center}
\renewcommand{\arraystretch}{1.2}
\setlength{\tabcolsep}{12pt}
\caption{The properties of the scalar quark-antiquark scalar states above $1\,{\rm GeV}$ and their decay modes.}
\label{table:meson information}
\begin{tabular}{c l l c c c}
\hline
&$J^{PC}$ & Decay Modes &${\cal B}(\%)$ & $\rm{Mass}$&$\rm{Width}$\\
\hline
$S$ &$$ &$M_{1}M_{2}$ &$S\to M_{1}M_{2}$ &${\rm MeV}$ &${\rm MeV}$
\\\hline
$f_0(1500)$ &$0^{ + + }$   &$\pi\pi/K\bar{K}$                  &$(34.5 \pm 2.2)/(8.5 \pm 1.0)$  &1522$\pm25$  &$108\pm33$ \\
$f_0(1370)$  &$0^{ + + }$ &$\pi\pi/\eta\eta$          &$\rm seen$ &$1200-1500$  &$200-500$ \\
$a_0(1450)$  &$0^{ + + }$ &$\gamma\gamma$      &$\rm seen$ &$1439\pm34$  &$258\pm14$  \\
$K_0^*(1430)$ & $0^ + $ & $K\pi$ & $(93 \pm 10)$ & $1425\pm50$ & $270\pm80$  \\
\hline
\end{tabular}
\end{center}
\end{table}

Currently, in scientific research, various decay channels such as semileptonic decays and rare decays are often used to explore the internal components of scalar resonances. Among them, the $B$ meson decays dominated by the flavor-changing neutral current (FCNC) $b\to s\ell^ + \ell^-(\ell=e,\mu,\tau)$ process have long been a key focus in particle physics research.
In the Standard Model (SM) of particle physics, this process cannot happen via the simplest, most direct ``tree-level'' interactions. Instead, it only occurs through indirect quantum effects called ``loop diagrams''. Its decay rate is strongly suppressed. The process relies on the CKM matrix element $|V_{tb}V_{ts}^*|$ a parameter in the theory that describes quark flavor changes~\cite{Kobayashi:1973fv}, which has a very small value. One the other hand, processes $b\to s\ell^ + \ell^-(\nu\bar\nu)$ are further suppressed by the GIM mechanism, a quantum effect that causes different particle contributions to cancel each other out. Because of these multiple suppressions, the predicted rate of these FCNC decays in the SM is extremely low (typically on the order of $10^{-6}$ or less). This makes them ``sensitive probes'' for testing the SM: The measurable properties of these decays (like angular distributions, lepton asymmetries, and lepton flavor universality (LFU) ratios~\cite{Hurth:2014vma, Das:2018sms,LHCb:2013ghj, Huber:2023qse, Lunghi:2010tr, Mahata:2022cxf, Falahati:2015hwa}) are highly sensitive to the effects of ``virtual particles'' predicted by theories beyond the SM. When this kind of FCNC rare decay of the $B$-meson develops into a quasi-four-body final state, the reaction can pass through an intermediate scalar resonance $f_0(1500)$. Afterwards, $f_0(1500)$ further decays into the two-body hadron channels $\pi\pi$ and $K\bar{K}$. Their corresponding branching fractions are $(34.5\pm2.2)\%$ and $(8.5\pm1.0)\%$, respectively. At the same time, calculating this quasi-four-body rare-decay process with $f_0(1500)$ helps us to describe fundamental particles and the inner structure of scalar resonances more accurately. Both experimentally and theoretically, it is of great importance to test the SM and search for possible new physics beyond it.

The main task of investigating the quasi-four-body rare decays $B_s \to f_0(1500)(\to \pi^ + \pi^-)\ell^ + \ell^-$ is to properly evaluate the hadronic matrix elements for $B_s \to f_0(1500)$ transition, namely the TFFs (most notably the vector and tensor form factors that dominate the rare decay amplitudes), which are governed by the non-perturbative QCD dynamics and act as the dominant source of theoretical uncertainty for the phenomenological predictions. There are several methods to solve this problem in the literature, such as simple quark model~\cite{Wirbel:1985ji}, light-front approach~\cite{Cheung:1995ub,Zhang:1994hg,Choi:1999nu}, QCD sum rules (QCDSR)~\cite{Shifman:1978by,Novikov:1981xi}, light-cone QCD sum rules (LCSR)~\cite{Braun:1988qv,Chernyak:1990ag}, perturbative QCD factorization approach~\cite{Keum:2000ph,Keum:2000wi,Lu:2000em} . The QCDSR approach is a fully relativistic approach and has made a tremendous success in hadron physics; however, short distance expansion fails in non-perturbative condensate when applying the three-point sum rules to the computations of $B_s \to f_0(1500)$ form factors in the large momentum transfer or large mass limit of heavy meson decays, which is a critical kinematic region for the experimental measurements and new physics searches of $B_s \to f_0(1500)(\to \pi^ + \pi^-)\ell^ + \ell^-$ quasi-four-body rare decays. LCSR is a natural combination of SVZ sum rules (SVZSR) and theory of hard exclusive process~\cite{Lepage:1979zb,Lepage:1980fj,Brodsky:1981rp,Efremov:1979qk,Efremov:1978rn,Chernyak:1983ej}, and has been widely tested and applied in theoretical studies of heavy-to-light hadron transitions~\cite{Cheng:2017bzz, Tian:2023vbh, Gao:2019lta, Duplancic:2008ix}. Compared with traditional SVZSR, LCSR uses LCDAs to describe nonperturbative effects, this allows for a quantitative analysis of high-twist non-perturbative corrections, which is difficult to achieve with traditional methods. At the same time, its calculation framework only involves a single Borel transformation and dispersion relation, which greatly improves computational efficiency. For these reasons, we use the LCSR method in this work to study the quasi-four-body rare decay $B_s \to f_0(1500)(\to \pi^ + \pi^-)\ell^ + \ell^-$. Furthermore, unlike the earlier LCSR analysis in Ref.~\cite{Huang:2021owr}, our calculation of the TFFs includes contributions from both twist-2 and twist-3 LCDA of the $f_0(1500)$ state. Meanwhile, phenomenologically, LCSR has been widely applied to investigate the rare semileptonic decays of heavy hadrons~\cite{Ball:1998tj, Khodjamirian:2000ds, Duplancic:2008ix, Wang:2007fs}, radiative hadronic decays~\cite{ Ali:1993vd, Aliev:1995zlh, Wang:2008sm}, non-leptonic two body rare decays of $B_s$-mesons~\cite{Khodjamirian:2000mi, Khodjamirian:2002pk, Khodjamirian:2003eq, Khodjamirian:2005wn} and strong coupling constants~\cite{Belyaev:1994zk}, and it has become one of the mainstream and reliable theoretical tools for the study of quasi-four-body rare decay.

In the heavy to light transition for the quasi-four-body rare decay $B_s \to f_0(1500)(\to \pi^ + \pi^-)\ell^ + \ell^-$, the twist-2 LCDA of the $f_0(1500)$-state is a key nonperturbative parameter. It not only directly determines the behavior and calculation accuracy of the TFFs, but also carries information about long-range QCD dynamics at low energy scales. For this reason, accurately determining the behavior of its twist-2 LCDA has long been a major focus in the field.
Traditionally, the twist-2 LCDA of the $f_0(1500)$-state is usually described by a truncated Gegenbauer series that keeps only the first few terms~\cite{Cheng:2005nb,LatticeParton:2022zqc}, and its coefficients known as Gegenbauer moments, can be calculated using QCD sum rules. To further reveal the internal dynamics of $f_0(1500)$-state from a phenomenological perspective, we can also construct its twist-2 LCDA using the LCHO model. This model is based on the Brodsky-Huang-Lepage (BHL) prescription and includes Wigner-Melosh rotations, which together form its theoretical framework. It also connects equal-time wave functions in the rest frame with light-cone wave functions in the infinite-momentum frame, and converts them into a relativistic form in light-cone coordinates. The LCHO model can describe both the spatial and spin components of the wave function, so it effectively characterizes the momentum distribution inside the meson~\cite{Huang:1994dy}, making it suitable for calculating $B_s \to f_0(1500)$ TFFs.
On this basis, we introduce the longitudinal correction function $\varphi_{2;f_0(1500)}(x)$ into the twist-2 wave function and construct an independent twist-2 LCDA scheme for the $f_0(1500)$ state. By comparing the calculated experimental observables for the quasi-four-body rare decay $B_s \to f_0(1500)(\to \pi^ + \pi^-)\ell^ + \ell^-$ under this scheme, we can not only test the SM of particle physics, but also verify its consistency and reliability.

In the present work. In Section~\ref{sec:II}, we present the decay-width formula of the quasi-four-body rare decay $B_s \to f_0(1500)(\to\pi^ + \pi^-)\ell^ + \ell^-$. Then we introduce the calculation of TFFs within the framework of the LCSR approach. Based on the LCHO model, we construct the twist-2 LCDA scheme for $f_0(1500)$. Subsequently, we calculate its moments $\langle \xi^n \rangle \big|_\mu$ and the Gegenbauer moments $a_n(\mu)$. Section~\ref{sec:III}, we show our numerical results on TFFs, which are further used to calculate the differential decay widths and branching ratios of this quasi-four-body rare decay. In Section~\ref{Sec:IV}, is used to be a summary.

\section{Theoretical Framework}\label {sec:II}
The tree-level interactions in the Standard Model relevant to the $b \to s\ell^ + \ell^-$ transition are described by the following Lagrangians:
\begin{align}
\mathcal{L}_{W} &= -\frac{g}{\sqrt{2}} \left( \bar{u}_i \gamma^\mu \frac{1-\gamma^5}{2} V_{ij} d_j \right) W_\mu^ +  + \text{h.c.}, \nonumber\\
\mathcal{L}_{Z} &= -\frac{g}{2\cos\theta_W} \sum_f \bar{f} \gamma^\mu \left( g_V^f - g_A^f \gamma^5 \right) f \, Z_\mu, \nonumber\\
\mathcal{L}_{\gamma} &= -e \sum_f Q_f \bar{f} \gamma^\mu f \, A_\mu.
\end{align}
Here $g$ is the $SU(2)_L$ gauge coupling constant, $\theta_W$ is the weak mixing angle, $V_{ij}$ are the CKM matrix elements, $Q_f$ denotes the electric charge of fermion $f$ in units of $e$, with $g_V^f = T_3^f/2 - Q_f \sin^2\theta_W$ and $g_A^f = T_3^f/2$. At tree level, there is no FCNC coupling in the SM; the $b \to s\ell^ + \ell^-$ transition is strictly forbidden by the GIM mechanism and can only occur through one-loop penguin diagrams (photon penguin and $Z$ penguin) and box diagrams.

Taking the photon penguin diagram as an example, its amplitude is proportional to $\bar s \gamma_\mu (1-\gamma^5)b \cdot \bar{\ell}\gamma^\mu \ell$, arising from the charged-current vertex in $\mathcal{L}_{W}$ and the electromagnetic vertex in $\mathcal{L}_{\gamma}$. The $Z$ penguin diagram involves the neutral-current coupling in $\mathcal{L}_{Z}$, while the box diagrams originate from two charged-current vertices of $\mathcal{L}_{W}$. These loop diagrams contain heavy internal particles, including $W^\pm$, $Z$, and $t$. In the low-energy limit $p \ll M_W, M_Z, m_t$, the propagators of these heavy particles can be expanded in Taylor series. Taking the $W$ propagator as an example:
\begin{align}
\frac{1}{k^2 - M_W^2} = -\frac{1}{M_W^2} \left( 1 + \frac{k^2}{M_W^2} + \frac{k^4}{M_W^4} + \cdots \right).
\end{align}
By retaining only the leading term, the heavy-particle effects reduce to local interaction vertices. Integrating out all heavy degrees of freedom ($W^\pm$, $Z$, $t$) via path integral:
\begin{align}
&\int \mathcal{D}[W^\pm,Z,t]\,e^{iS_{\text{SM}}} \longrightarrow \exp\left[iS_{\text{eff}}\right],\nonumber\\
&\mathcal{S}_{\text{eff}} \equiv \int d^4x\,\mathcal{H}_{\text{eff}}(x)
\end{align}
the loop contributions are absorbed into the Wilson coefficients $C_i(\mu)$ multiplying a set of local operators $\mathcal{O}_i$ constructed from the light fields ($b, s, \ell$). The matching condition requires that at the scale $\mu = M_W$, the effective theory amplitude equals the full Standard Model amplitude, $\mathcal{A}_{\text{SM}}^{\text{full}} = \mathcal{A}_{\text{eff}}$.
Subsequently, the Wilson coefficients are evolved from $\mu = M_W$ down to $\mu = m_b$ via the renormalization group equations, $\mu \frac{d}{d\mu} C_i(\mu) = \gamma_{ji} \, C_j(\mu)$,
where $\gamma_{ji}$ is the anomalous dimension matrix. This evolution resums the large logarithms $\ln(M_W/m_b)$ to all orders. Consequently, the low-energy effective Hamiltonian for the $b \to s\ell^ + \ell^-$ transition is obtained as
\begin{align}
\mathcal{H}_{\text{eff}}(b \to s\ell^ + \ell^-) = -\frac{4G_F V_{tb} V_{ts}^*}{\sqrt{2}} \sum_{i=1}^{10} C_i(\mu) \mathcal{O}_i(\mu),
\end{align}
where $G_{\rm F}$ is the Fermi constant, with $G_F/\sqrt{2} = g^2/(8M_W^2)$ , $V_{tb}V_{ts}^*$ being the CKM factor and $C_i$ are Wilson coefficients. The operators $\mathcal{O}_i(\mu)$ $(i = 1, \ldots, 10)$ are classified according to their physical nature into four-quark operators $(i = 1, \ldots, 6)$, the electromagnetic dipole operator $(i = 7)$, the gluonic dipole operator $(i = 8)$, and the semileptonic operators $(i = 9, 10)$. These operators provide the main contribution in SM and the explicit forms for the decay $B_s\to f_0(1500)\ell^ + \ell^-$ are~\cite{Tian:2024ubt}
\begin{align}
&{\cal O}_1 = (\bar s _{\alpha} \gamma^{\mu} (1-\gamma_5) b_\alpha)(\bar{c}_\beta \gamma_{\mu}(1-\gamma_5) c_\beta),\nonumber\\
&{\cal O}_2 = (\bar s _{\alpha} \gamma^{\mu} (1-\gamma_5) b_\beta)(\bar{c}_\beta \gamma_{\mu}(1-\gamma_5) c_\alpha),\nonumber\\
&{\cal O}_3 = (\bar s _\alpha \gamma^\mu (1 - \gamma_5) b_\alpha) \sum_{q=u,d,s,c,b} (\bar{q}_\beta \gamma_\mu (1 - \gamma_5) q_\beta),\nonumber\\
&{\cal O}_4 = (\bar s _\alpha \gamma^\mu (1 - \gamma_5) b_\beta) \sum_{q=u,d,s,c,b} (\bar{q}_\beta \gamma_\mu (1 - \gamma_5) q_\alpha),\nonumber\\
&{\cal O}_5 = (\bar s _\alpha \gamma^\mu (1 - \gamma_5) b_\alpha) \sum_{q=u,d,s,c,b} (\bar{q}_\beta \gamma_\mu (1 + \gamma_5) q_\beta),\nonumber\\
&{\cal O}_6 = (\bar s _\alpha \gamma^\mu (1 - \gamma_5) b_\beta) \sum_{q=u,d,s,c,b} (\bar{q}_\beta \gamma_\mu (1 + \gamma_5) q_\alpha),\nonumber\\
&{\cal O}_7 =\frac{e}{16 {\pi^2}}m_b (\bar s _\alpha \sigma^{\mu \nu}(1 + \gamma_5) b_{\alpha})F_{\mu \nu},\nonumber\\
&{\cal O}_8 =\frac{g_s}{16\pi^2} m_b \left( \bar s _\alpha \sigma^{\mu\nu} (1 + \gamma_5) T^a_{\alpha\beta} b_\beta \right) G^a_{\mu\nu},\nonumber\\
&{\cal O}_9 =\frac{e^2}{16 {\pi^2}}(\bar s _\alpha \gamma^{\mu}(1-\gamma_5) b_\alpha)\bar{\ell}\gamma_{\mu}\ell ,\nonumber\\
&{\cal O}_{10} =\frac{e^2}{16 {\pi^2}} (\bar s _\alpha \gamma^{\mu} (1-\gamma_5)b_\alpha)\bar{\ell}\gamma_{\mu}\gamma_5 \ell.
\label{eq.O1,10}
\end{align}
Here, $g_s$ is the strong coupling constant, $F_{\mu\nu}$ is the electromagnetic field strength tensors, $\alpha$ and $\beta$ are colour indices, $\sigma^{\mu\nu}= \frac{i}{2}[\gamma^{\mu},\gamma^{\nu}]$, while the sum runs over all the five active quark flavors, i.e. $p = (u, d, s, c, b)$. At the $b$-quark scale $\mu \sim m_b$, the top quark has been integrated out and is no longer a dynamical field. This leaves five active dynamical quarks: $u$, $d$, $s$, $c$, and $b$. In penguin-type loop diagrams, virtual gluons can produce quark-antiquark pairs of these five flavours. Therefore, the penguin operators carry an explicit sum over these quark flavours. All loop-level effects originating from the top quark are encoded within the Wilson coefficients, rather than appearing in the flavour summation of the operators. Where the coefficients of the QCD penguin operators ${\cal O}_{3-6}$ are small. The reasons are as follows. Firstly, they are nearly zero at the electroweak scale, and are only generated very weakly through renormalization group mixing.
Secondly, their matrix elements are further suppressed by the chirality (handedness) symmetry.
Finally, Compared with electromagnetic or semileptonic operators, their numerical contributions are higher-order small quantities.
Therefore, in most phenomenological analyses of the SM, their contributions can be neglected unless extremely high precision is required (such as NNLO or higher-order perturbative calculations).

In the SM, the $b \to s \nu \bar{\nu}$ process is forbidden at tree level and proceeds only through FCNC transitions at the one-loop level. To describe this process, we employ the operator product expansion (OPE) approach, integrating out the heavy degrees of freedom above the electroweak scale $M_W$, thereby obtaining the low-energy effective Hamiltonian applicable at the $b$-quark scale $\mu_b \sim \mathcal{O}(m_b)$. The electroweak interactions governing the $b \to s \nu \bar{\nu}$ transition originate from the SU$(2)_L \times {\rm U}(1)_Y$ gauge theory. The charged-current interaction Lagrangian is given by
\begin{align}
\mathcal{L}_{\rm CC} = -\frac{g}{\sqrt{2}} \left[ \bar{u}_i \gamma^\mu P_L d_i W_\mu^ +  + \bar{\nu}_\ell \gamma^\mu P_L \ell W_\mu^ +  + \text{h.c.} \right]
\label{eq:Lc}
\end{align}
while the neutral-current interaction takes the form $\mathcal{L}_{NC} = -\frac{g}{2\cos\theta_W} \sum_f \bar{f} \gamma^\mu (g_V^f - g_A^f \gamma_5) f Z_\mu$,
where $P_L = (1 - \gamma_5)/2$, $g$ is the SU(2)$_L$ coupling constant, and $\theta_W$ is the weak mixing angle. For neutrinos, with $T_3^\nu = + 1/2$ and $Q_\nu = 0$, we have $g_V^\nu = g_A^\nu = 1/2$, leading to a purely left-handed coupling between the $Z$ boson and neutrinos, $\mathcal{L}_{NC}^{\nu} = -\frac{g}{2\cos\theta_W} \bar{\nu} \gamma^\mu P_L \nu Z_\mu$.
For the $b \to s$ flavor-changing neutral current, the tree-level $Z$ coupling vanishes in the SM due to the GIM mechanism, and the effective vertex is generated entirely at the loop level.

At low energies $\mu \ll M_W$, the heavy degrees of freedom, namely the $W^\pm$, $Z$-bosons, and the top quark, are integrated out. At the one-loop level, the process receives two types of contributions: the $W$-box diagrams and the $Z$-penguin diagrams. A direct matching calculation, performed using dimensional regularization ($d = 4 - 2\epsilon$) and the $\overline{\rm MS}$ renormalization scheme, yields the Wilson coefficients for both types of diagrams. After Dirac algebra reduction, the quark current part reduces to the current-current structure $(\bar s \gamma_\mu P_L b)(\bar{\nu} \gamma^\mu P_L \nu)$ in the chiral limit. Summing the box and penguin contributions, the total Wilson coefficient is obtained as
\begin{align}
C_{\rm total} = \frac{e^2}{16\pi^2} \frac{x_t}{\sin^2\theta_W} \left[
\frac{3(x_t - 2)}{2(x_t - 1)} + \frac{9x_t - 2}{2(x_t - 1)^2} \ln x_t \right],
\label{eq:Ctotal}
\end{align}
where $x_t = m_t^2 / M_W^2$ and $e = g \sin\theta_W$ is the elementary charge. This result can be compactly expressed in terms of the standard one-loop function $X(x_t)$, defined as~\cite{Inami:1980fz}
\begin{align}
X(x_t) = \frac{x_t}{8} \left[ \frac{x_t + 2}{x_t - 1} + \frac{3x_t - 6}{(x_t - 1)^2} \ln x_t \right],
\label{eq:Xfunction}
\end{align}
which fully captures the dependence on the top-quark mass. Substituting Eq.~(\ref{eq:Xfunction}) into Eq.~(\ref{eq:Ctotal}), one verifies that $C_{\rm total} = 4e^2X(x_t)/(16\pi^2 \sin^2\theta_W)$.

For the double neutrinos final state in $b \to s \nu \bar{\nu}$ decay process are left-handed in the SM, and $b \to s$ current is purely left-handed. We define
\begin{align}
O_L &= \frac{e^2}{16\pi^2} (\bar s \gamma_\mu (1 - \gamma_5) b)(\bar{\nu} \gamma^\mu (1 - \gamma_5) \nu),
\label{eq:OL}
\\
O_R &= \frac{e^2}{16\pi^2} (\bar s \gamma_\mu (1 + \gamma_5) b)(\bar{\nu} \gamma^\mu (1 - \gamma_5) \nu).
\label{eq:OR}
\end{align}
The low-energy effective Hamiltonian can then be uniformly written as
\begin{align}
H_{\text{eff}}(b \to s \nu \bar{\nu}) = -\frac{4G_F}{\sqrt{2}} V_{tb} V_{ts}^* ( C_L^{\rm SM} O_L + C_R^{\rm SM} O_R )
\label{eq:Heff}
\end{align}
where $G_F$ is the Fermi constant satisfying $G_F/\sqrt{2} = g^2/(8M_W^2)$, and $V_{tb}V_{ts}^*$ is the corresponding CKM matrix element production. From the matching calculation above, the SM Wilson coefficients are given by
\begin{align}
C_L^{\text{SM}} = \frac{X(x_t)}{\sin^2\theta_W}, \qquad C_R^{\text{SM}} = 0.
\label{eq:CLCR}
\end{align}
The result $C_R^{\text{SM}} = 0$ in Eq.~(\ref{eq:CLCR}) is exact rather than approximate. Its physical origin is as follows: the $W^\pm$ bosons couple only to left-handed currents in the SM, and the $Z$ boson coupling to neutrinos is also purely left-handed. To generate a right-handed current structure of the form $\bar s \gamma_\mu (1 + \gamma_5) b$, one would need to simultaneously flip the chirality of both the quark and lepton lines, which corresponds to an amplitude proportional to $m_t m_\nu$. Since neutrinos are strictly massless in the SM ($m_\nu = 0$), this contribution vanishes exactly. Consequently, any deviation from $C_R = 0$ would constitute a clean signature of new physics beyond the SM. Then, the free quark amplitude for $b\to s\ell^ + \ell^-$ and $b\to s\nu\bar\nu$ can be derived as~\cite{Chen:2007na}
\begin{align}
&\mathcal{M}(b\to s\ell^ + \ell^-) = \langle s \ell^ + \ell^-| \mathcal{H}_{\rm eff} |b \rangle
\nonumber
\\
&\hspace{2.5cm} = \frac{G_F \alpha_{\rm em}}{\sqrt{2}\pi} V_{ts}^\ast V_{tb} \Big[C_9^{\rm eff}(m_b) \left( \bar s \gamma_{\mu} (1-\gamma_5)b\bar{\ell}\gamma^{\mu}\ell \right) + C_{10}( \bar s \gamma_{\mu} (1-\gamma_5)b\bar{\ell}\gamma^{\mu}\gamma_5\ell )
\nonumber
\\
&\hspace{2.5cm}- 2iC_7^{\rm eff}(m_b) \frac{m_b }{q^2} \left(\bar s \sigma_{\mu\nu}q^{\nu}(1 + \gamma_5)b\bar{\ell}\gamma^{\mu}\ell \right) \Big],
\label{Eq:Heff-bsll}
\\
&\mathcal{M}(b\to s\nu\bar{\nu}) = \langle s \nu\bar{\nu}| \mathcal{H}_{\rm eff} |b \rangle
\nonumber
\\
&\hspace{2.2cm}
= \frac{G_F \alpha_{\rm em}}{2\pi\sqrt{2} \sin^2\theta_w} V_{ts}^\ast V_{tb}X_0(x_t) \bar{b}\gamma_{\mu}(1-\gamma_5)s\bar{\nu}_\ell \gamma_{\mu}(1-\gamma_5)\nu_\ell,
\label{Eq:Heff-bsvv}
\end{align}
where $\alpha_{\rm em}$ is the fine structure constant at $Z$-boson mass scale. The $C_{10}$ is independent on the energy scale, because there is no $Z$-boson in the effective theory, the operator ${\cal O}_{10}$ cannot be introduced by the insertion of four-quark operators~\cite{Wang:2008da}. In addition, the corresponding quark decay amplitude can receive short-distance and long-distance contributions from the matrix element of current operators ${\cal O}_1$ and ${\cal O}_2$, which can be quantified into the effective Wilson coefficient $C_9^{\rm eff}(m_b)$~\cite{Maji:2018gvz}. $C_9^{\rm eff}(m_b)= C_9 + Y(\hat{s})$, where $Y(\hat{s}) = Y_{\rm pert}(\hat{s}) + Y_{\rm LD}$ contains both the perturbative part $Y_{\rm pert}(\hat{s})$ and long-distance part $Y_{\rm LD}$ from the four quark operators~\cite{Buras:1994dj,Buchalla:1995vs}
\begin{align}
Y_{\rm pert}(\hat{s}) &= h(z,\hat{s})(3C_1 + C_2 + 3C_3 + C_4 + 3C_5 + C_6)
- \frac{1}{2}h(1,\hat{s})(4C_3 + 4C_4 + 3C_5 + C_6)\nonumber\\
& - \frac{1}{2}h(0,\hat{s})(C_3 + 3C_4)
 + \frac{2}{9}(3C_3 + C_4 + 3C_5 + C_6),
\label{Eq:Ypert}
\end{align}
with
\begin{align}
h(z,\hat{s})&= -\frac{8}{9}\ln z + \frac{8}{27} + \frac{4}{9}x - \frac{2}{9}(2 + x)|1-x|^{1/2}\nonumber\\
&\times \begin{cases}
&\bigg|\dfrac{\sqrt{1-x} + 1}{\sqrt{1-x} - 1} \bigg| - i\pi, x\equiv 4z^2/\hat{s} < 1
\\
&2\arctan\dfrac{1}{\sqrt{x-1}}, x\equiv 4z^2/\hat{s} > 1
\end{cases}
\end{align}
and the $h(0,\hat{s})$ have the expression
\begin{align}
h(0,\hat{s})=\frac{8}{27} - \frac{8}{9}\ln \frac{m_b}{\mu} - \frac{4}{9}\ln\hat{s} + \frac{4}{9}i\pi,
\end{align}
where $z= m_c/m_b$ and $\hat{s}=q^2/m_b^2$.
Nevertheless, $f_0(1500)$ is an unstable scalar resonance that couples strongly to both the $\pi^ + \pi^-$ and $K\bar{K}$ decay channels. In the full quasi-four-body decay $B_s \to f_0(1500)(\to\pi^ + \pi^-)\ell^ + \ell^-$, the intermediate resonance can be off-shell, so the fixed-mass approximation is no longer valid. Within the intermediate-resonance approximation, the full quasi-four-body decay amplitude can be factorized into three components: the weak decay amplitude for the three-body rare decay, the resonance propagator, and the strong-decay vertex for $f_0(1500)\to\pi^ + \pi^-$~\cite{Wang:2016wpc};
\begin{align}
\mathcal{A}\big(B_s\to \pi^ + \pi^-\ell^ + \ell^-\big)=\hat{\mathcal{A}}\cdot\left(\frac{i}{D_{f_0(1500)}(s)}\times i\,g_{f_0(1500)\pi^ + \pi^-}\right),
\label{eq:Amp_total}
\end{align}
where $\hat{\mathcal{A}}$ denotes the weak amplitude of the three-body rare decay $B_s\to f_0(1500)\ell^ + \ell^-$ treating $f_0(1500)$ as a stable particle, $g_{f_0(1500)\pi^ + \pi^-}$ is the strong coupling for $f_0(1500)\to\pi^ + \pi^-$.
The core component of the hadronic decay amplitude $\hat{A}$ is the weak transition matrix element for the $B_s \to f_0(1500)$ transition, whose Lorentz structure can be fully parameterized in terms of the TFFs.
\begin{align}
\langle f_0(1500)(p) | i\bar s \gamma_\mu \gamma_5 b | B_s(p + q) \rangle
&= 2 f_ + ^{B_s f_0(1500)}(q^2) p_\mu
 + \Big[ f_ + ^{B_s f_0(1500)}(q^2)
\nonumber
\\
& + f_-^{B_s f_0(1500)}(q^2) \Big] q_\mu,
\label{eq:aa}
\\
\langle f_0(1500)(p) | \bar s \sigma_{\mu\nu} \gamma_5 q^\nu b | B_s(p + q) \rangle
&= \frac{f_{\rm T}^{B_s f_0(1500)}(q^2)}{m_{B_s} + m_{f_0(1500)}} [ 2p_\mu q^2 - 2q_\mu (p \cdot q) ] .
\label{eq:bb}
\end{align}
Here, the axial-vector current matrix element is described by $f_ + ^{B_s f_0(1500)}(q^2)$ and $f_-^{B_s f_0(1500)}(q^2)$, while the tensor current matrix element is characterized by $f_{\rm T}^{B_s f_0(1500)}(q^2)$. When combining $\hat{A}$ with the strong coupling constant $g_{f_0(1500)\pi^ + \pi^-}$, the hadronic part of the decay amplitude is fully determined by these TFFs.
The quantity $D_{f_0(1500)}(s)$ is the denominator of the resummed resonance propagator, which incorporates self-energy corrections, finite-width effects and coupling information of $f_0(1500)$:
\begin{align}
D_{f_0(1500)}(s)=m_{f_0(1500)}^2-s + \Pi_{f_0(1500)}(s).
\label{eq:Df0_def}
\end{align}
Here $\Pi_{f_0(1500)}(s)$ stands for the self-energy function. Its real part shifts the pole position of the propagator and accounts for the correction from bare mass to physical mass of $f_0(1500)$; its imaginary part is directly related to the energy-dependent partial decay widths and characterizes the probability for $f_0(1500)$ to decay strongly into $\pi\pi$ and $K\bar{K}$ channels. One can further write~\cite{Flatte:1976xv,LHCb:2014ooi}
\begin{align}
D_{f_0(1500)}(s)=m_{f_0(1500)}^2-s-i\sqrt{s}[\Gamma_{f_0(1500)\to\pi^ + \pi^-}(s) + \Gamma_{f_0(1500) \to K^ + K^-}(s)].
\label{eq:Df0}
\end{align}
$\Gamma_{f_0(1500)\to\pi^ + \pi^-}(s)$ and $\Gamma_{f_0(1500)\to K^ + K^-}(s)$ are nonzero mass-dependent decay width, which the specific expressions are respectively:
\begin{align}
&\Gamma_{f_0(1500)\to\pi\pi}(s)=\frac{1}{16\pi\sqrt{s}}|g_{f_0(1500)\pi^ + \pi^-}|^2\rho_{\pi\pi}(s),\\
&\Gamma_{f_0(1500)\to K\bar K}(s)=\frac{1}{16\pi\sqrt{s}}|g_{f_0(1500)K^ + K^-}|^2\rho_{K\bar K}(s).
\label{eq:partial_width}
\end{align}
where $g_{f_0(1500)\pi^ + \pi^-}=\sqrt{16\pi g_1}$ and $g_{f_0(1500)\ K^ + K^-}=\sqrt{16\pi g_2}$, with the constants $g_1$ and $g_2$ are the $f_0(1500)$ couplings to $\pi^ + \pi^-$ and $K^ + K^-$ final states, respectively. The factors $\rho_{\pi\pi}$ and $\rho_{K{\bar K}}$ are given by Lorentz invariant phase space~\cite{Achasov:2020qfx}
\begin{align}
&\rho_{\pi\pi}(s)=\sqrt{\left[1-\frac{(m_\pi-m_\pi)^2}{s}\right]\left[1-\frac{(m_\pi + m_\pi)^2}{s}\right]},
\\
&\rho_{K{\bar K}}(s)=\sqrt{\left[1-\frac{(m_K-m_{\bar K})^2}{s}\right]\left[1-\frac{(m_K + m_{\bar K})^2}{s}\right]}.
\end{align}
We adopt the relativistic Flatt\'e line-shape function to describe the invariant-mass distribution of $f_0(1500)$, which simultaneously includes the open $\pi\pi$ and $K\bar{K}$ channels:
\begin{align}
\hspace{-0.2cm} P(s)=\frac{g_1\rho_{\pi\pi}}{|m^2_{f_0(1500)} - s - i[g_1 \rho_{\pi\pi} (s) + g_2\rho_{K{\bar K}}(s)]|^2}.
\label{eq:flatte_f01500}
\end{align}
Here the $\rho_{\pi\pi}$ and $\rho_{K{\bar K}}$ are individual phase space factors. The normalization condition satisfies $\int_{(2m_\pi)^2}^{\infty} ds\, P(s)/\pi=1$. After integrating over the angular phase space of the $\pi^ + \pi^-$ system, the resonance effect is implemented by multiplying the three-body decay spectrum with the weight factor $P(s)/\pi$.
Averaging over the initial-state spin of $B_s$ and summing over final-state lepton spins yields the single-differential decay width $d\Gamma/dq^2$ for the three-body process $B_s\to f_0(1500)\ell^ + \ell^-$ under the stable-resonance assumption, where $s=m_{f_0(1500)}^2$ is fixed to the nominal resonance mass. In this approximation,
$s$ is not an independent variable but is fixed to the resonance pole and all hadronic-leptonic dynamics are encoded in $|\hat{\mathcal A}|^2$. The double differential decay width for the quasi-four-body decay process can be written as:
\begin{align}
&\frac{d\Gamma(B_s \to f_0(1500)(\to \pi^ + \pi^-)\ell^ + \ell^-)}{ds\,dq^2}
= \frac{G_F^2 |V_{tb}V_{ts}^*|^2 m_{B_s}^3 \alpha_{\mathrm{em}}^2}{1536\pi^5}
\sqrt{1 - \frac{4m_\ell^2}{q^2}}
\nonumber\\
&\hspace{2cm}
\times \Bigg\{ \bigg(1 + \frac{2m_\ell^2}{q^2}\bigg)
\Bigg[ \bigg(1 - \frac{s}{m_{B_s}^2}\bigg)^2
- \frac{2 q^2}{m_{B_s}^2}\bigg(1 + \frac{s}{m_{B_s}^2}\bigg)
 + \bigg(\frac{q^2}{m_{B_s}^2}\bigg)^2 \Bigg]^{3/2}
\nonumber\\
&\hspace{2cm}
\times \bigg[ \bigg| C_9^{\mathrm{eff}} f_ + ^{B_s f_0(1500)}(q^2)
- \frac{2m_{B_s}}{m_{B_s} + \sqrt{s}}\, C_7^{\mathrm{eff}} f_{\rm T}^{B_s f_0(1500)}(q^2) \bigg|^2
 + \big| C_{10} f_ + ^{B_s f_0(1500)}(q^2) \big|^2 \bigg]
\nonumber\\
&\hspace{2cm}
 + \Bigg[ \bigg(1 - \frac{s}{m_{B_s}^2}\bigg)^2
- \frac{2 q^2}{m_{B_s}^2}\bigg(1 + \frac{s}{m_{B_s}^2}\bigg)
 + \bigg(\frac{q^2}{m_{B_s}^2}\bigg)^2 \Bigg]^{1/2}
\frac{m_\ell^2}{m_{B_s}^2} 6\,|C_{10}|^2
\nonumber\\
&\hspace{2cm}
\times \Bigg\{ \Bigg[ 2\bigg(1 + \frac{s}{m_{B_s}^2}\bigg)
- \frac{q^2}{m_{B_s}^2} \Bigg] \big| f_ + ^{B_s f_0(1500)}(q^2) \big|^2
\nonumber\\
&\hspace{2cm}
 + 2\bigg(1 - \frac{s}{m_{B_s}^2}\bigg)
\operatorname{Re}\!\left[ f_ + ^{B_s f_0(1500)}(q^2) f_-^{B_s f_0(1500)}(q^2) \right]
\nonumber\\
&\hspace{2cm}
 + \frac{q^2}{m_{B_s}^2} \big| f_-^{B_s f_0(1500)}(q^2) \big|^2 \Bigg\} \Bigg\}\frac{P(s)}{\pi}.
\label{eq:DW1}
\end{align}
\begin{align}
&\frac{d\Gamma(B_s \to f_0(1500)(\to \pi^ + \pi^-) \nu\bar{\nu})}{ds dq^2}
=\frac{G_F^2 |V_{tb} V_{ts}^\ast|^2 m_{B_s}^3 \alpha_{\rm em}^2}{256\pi^5} \frac{x_t^2 \Big|(x_t + 2)(x_t-1) + 3(x_t-2)\ln x_t\Big|^2}{64(x_t-1)^4 \sin^4\theta_W}
\nonumber\\
&\hspace{1cm}\times\Bigg[ \bigg(1 - \frac{s}{m_{B_s}^2}\bigg)^2
- \frac{2 q^2}{m_{B_s}^2}\bigg(1 + \frac{s}{m_{B_s}^2}\bigg)
 + \bigg(\frac{q^2}{m_{B_s}^2}\bigg)^2 \Bigg]^{3/2}
\big| f_ + ^{B_s f_0(1500)}(q^2) \big|^2
\frac{P(s)}{\pi}.
\label{eq:DW2}
\end{align}
$s$ is treated as a continuous variable representing the invariant-mass squared of the $\pi^ + \pi^-$ system, $s\in[(m_K + m_\pi)^2, \; (m_{B_s} - m_\ell)^2]$, lepton-pair invariant-mass squared: $q^2\in [4m_\ell^2, \big(m_{B_s}-\sqrt{s}\big)^2]$.
Integrating the double-differential decay width over the full physical phase-space yields the total decay width
\begin{align}
\Gamma=\int_{(m_K + m_\pi)^2}^{(m_{B_s} - m_\ell)^2} ds \int_{4m_\ell^2}^{(m_{B_s}-\sqrt{s})^2} dq^2\;
\frac{d^2\Gamma\big(B_s \to f_0(1500)(\to\pi^ + \pi^-)\ell^ + \ell^-\big)}{ds dq^2}.
\label{eq:Gamma_total}
\end{align}
Combined with the $B_s$ meson lifetime $\tau_{B_s}$, the corresponding branching fraction is given by
\begin{align}
{\cal B}(B_s \to f_0(1500)(\to\pi^ + \pi^-)\ell^ + \ell^-)=\Gamma\cdot \tau_{B_s}.
\label{eq:Br}
\end{align}

Then, we need to obtain the behavior of the TFFs for $B_s\to f_0(1500)$ by LCSR. We first follow the standard steps of sum rules to introduce the vacuum-to-$f_0(1500)$ correlation function, which is related to the TFFs $f^{B_s f_0(1500)}_{\pm, {\rm T}}(q^2)$, which is defined as
\begin{align}
\Pi_\mu(p,q) &= i\int d^4x e^{iq\cdot x} \langle f_0(1500)(p)| T\{\bar s (x)\gamma_\mu \gamma_5 b(x), \bar{b}(0)i\gamma_5 s(0)\} |0\rangle
\nonumber\\
&= F(q^2,(p + q)^2)p_\mu + \widetilde{F}(p^2,(p + q)^2)q_\mu
\nonumber\\
\widetilde{\Pi}_\mu(p,q) &= i\int d^4x e^{iq\cdot x} \langle f_0(1500)(p)| T\{\bar s (x) \sigma_{\mu\nu} \gamma_5 q^\nu b(x), \bar{b}(0) i\gamma_5 s(0) \} |0\rangle
\nonumber\\
&= F^{\rm T}(p^2,(p + q)^2) [p_\mu q^2 - q_\mu (p\cdot q)].
\label{eq:correlatorTFFs}
\end{align}
For the decay $B_s\to f_0(1500)$ in Eq.~\eqref{eq:correlatorTFFs}, the light quark is $s$, and the heavy quark is $b$. The $p$ is the $f_0(1500)$-resonance four momentum, $q$ and $(p + q)$ are transition momentum and the $B_s$-meson momentum, respectively.
We first calculate the correlator~\eqref{eq:correlatorTFFs} in QCD.
Insert a complete set of hadronic states between the currents in correlator~\eqref{eq:correlatorTFFs} to obtain the hadronic representations of the invariant amplitudes. In which, the TFFs $f^{B_s f_0(1500)}_{\pm,{\rm T}}(q^2)$ enters the correlator~\eqref{eq:correlatorTFFs} via the hadronic matrix elements for the interpolating currents indicating the weak transition of $b$ to $s$. They can be parameterized in terms of the TFFs $f^{B_s f_0(1500)}_{\pm,{\rm T}}(q^2)$ as Eqs.~\eqref{eq:aa} and ~\eqref{eq:bb}
Otherwise, the vacuum-to-meson matrix element for the interpolating current representing the $B_s$ channel can be given by
\begin{align}
\langle{B_s}|\bar{b} i\gamma_5 s |0\rangle = \frac{{m_{B_s}^2} f_{B_s}}{m_b + m_s}
\label{eq:DecayConstant}
\end{align}
with the heavy meson mass $m_{B_s}$ and decay constant $f_{B_s}$. Then the hadronic representations of the invariant amplitudes $F$, $\widetilde{F}$ and $F^{\rm T}$ can be written as
\begin{align}
&F_{\rm had}(p^2,(p + q)^2)= \frac{-2i m_{B_s}^2 f_{B_s} f^{B_s f_0(1500)}_ + (q^2)}{(m_b + m_s) [m_{B_s}^2 - (p + q)^2]} + \cdots,
\nonumber\\
&\widetilde{F}_{\rm had}(p^2,(p + q)^2)= \frac{-i {m_{B_s}^2} f_{B_s} [f^{B_s f_0(1500)}_ + (q^2) + f^{B_s f_0(1500)}_-(q^2)]}{(m_b + m_s) [m_{B_s}^2 - (p + q)^2]} + \cdots,
\nonumber\\
&F_{\rm had}^{\rm T}(p^2,(p + q)^2)= \frac{-2m_{B_s}^2 f_{B_s} f^{B_s f_0(1500)}_{\rm T}(q^2)}{(m_b + m_s) (m_{B_s} + m_{f_0(1500)}) [m_{B_s}^2 - (p + q)^2]} + \cdots,
\label{eq:had}
\end{align}
respectively. In Eq.~\eqref{eq:had}, the ground state heavy meson contributions have been isolated, and the ellipses indicate the contributions from the excited states, the continuum states and possible subtraction terms.

Without losing generality, we take the invariant amplitude $F(q^2, (p + q)^2)$ as an example to illustrate the subsequent calculation procedure. One can write a general dispersion relation for $F(q^2, (p + q)^2)$ and further apply the Borel transformation with respect to the momentum squared $(p + q)^2$ of the heavy meson~\cite{Belyaev:1993wp},
\begin{align}
F(q^2,M^2) = \int^\infty_{\cal T} \rho(q^2,s) e^{-s/M^2} ds
\label{eq:DispersionRelationTFFs}
\end{align}
with ${\cal T} = (m_b + m_s)^2$. In which, the spectral density is given by
\begin{align}
\rho(q^2,s) &= \frac{1}{\pi} {\rm Im} F_{\rm had}(q^2,s)
\nonumber\\
&= \delta(s-m_{B_s}^2) \frac{-2i m_{B_s}^2 f_{B_s} f^{B_s f_0(1500)}_ + (q^2)}{m_b + m_s} + \frac{1}{\pi} {\rm Im} F_{\rm QCD}(q^2,s) \theta(s - s_{B_s}),
\label{eq:SpectralDensity}
\end{align}
where the contributions of the excited states and continuum states in $F_{\rm had}(q^2,s)$ have been parameterized as ${\rm Im} F_{\rm QCD}(q^2,s)/\pi$ and been delimited by the effective threshold parameter $s_{B_s}$ with the quark-hadronic duality approximation, and the possible subtractions will be got rid of due to Borel transformation in Eq.~\eqref{eq:DispersionRelationTFFs}. Substituting Eq.~\eqref{eq:SpectralDensity} into Eq.~\eqref{eq:DispersionRelationTFFs}, one can get
\begin{align}
F(q^2,M^2) &= \frac{-2i m_{B_s}^2 f_{B_s} f^{B_s f_0(1500)}_ + (q^2)}{m_b + m_s} e^{-m_{B_s}^2/M^2} + \frac{1}{\pi} \int^\infty_{s_{B_s}} {\rm Im} F_{\rm QCD}(q^2, s) e^{-s/M^2} ds.
\label{eq:LCSRhad}
\end{align}
On the other hand, the invariant amplitude after Borel transformation can also be written as~\cite{Belyaev:1993wp}
\begin{align}
F(q^2,M^2) &= \frac{1}{\pi} \int^\infty_{\cal T} {\rm Im} F_{\rm QCD}(q^2, s) e^{-s/M^2} ds.
\label{eq:LCSRqcd}
\end{align}
Finally, equating Eqs.~\eqref{eq:LCSRhad} with~\eqref{eq:LCSRqcd}, the LCSR of TFF $f_{\pm, {\rm T}}(q^2)$ can be obtained as
\begin{align}
f_ + ^{B_s f_0(1500)}(q^2) &= (m_b + m_s) \frac{\bar{f}_{f_0(1500)} }{2m_{B_s}^2 f_{B_s}} e^{m_{B_s}^2/M^2}\!\! \int^{\tilde x_0}_{x_0}\! dx \exp \bigg[-\frac{m_b^2-\bar x q^2 + x\bar x m_{f_0(1500)}^2}{xM^2}\bigg]
\nonumber
\\
&\times \bigg\{ -m_b \frac{\phi_{2;f_0(1500)}(x)}{x} + m_{f_0(1500)} \phi_{3;f_0(1500)}^p(x) + m_{f_0(1500)} \bigg[ \frac{2}{x} + 4x m_b^2
\nonumber
\\
&\times \frac{m_{f_0(1500)}^2}{(m_b^2 - q^2 + \! x^2 m_{f_0(1500)}^2)^2} \! - \! \frac{m_b^2 + q^2 - x^2 m_{f_0(1500)}^2}{m_b^2 - q^2 + x^2m_{f_0(1500)}^2} \frac{d}{dx} \bigg] \frac{\phi_{3;f_0(1500)}^\sigma(x)}{6} \bigg\}.
\label{eq:LCSRTFFs1}
\\
f_+^{B_s f_0(1500)}(q^2) & + f_-^{B_s f_0(1500)}(q^2) \, = \, (m_b \, + \, m_s) ~ \frac{m_{f_0(1500)} ~ \bar{f}_{f_0(1500)} }{m_{B_s}^2 f_{B_s}} ~\, e^{\,m_{B_s}^2\,/\,M^2} ~\,
\int^{\tilde{x}_0}_{x_0} \, dx
\nonumber\\
&\times \exp\bigg[ \! -\frac{m_b^2 - \! \bar x q^2 + \! x\bar x m_{f_0(1500)}^2}{xM^2}\bigg]\bigg[ \frac{\phi_{3;f_0(1500)}^p(x)}{x}  + \! \frac{1}{6x} \frac{d}{dx} \phi_{3;f_0(1500)}^\sigma(x) \bigg]
\label{eq:LCSRTFFs2}
\\
f_{\rm T}^{B_s f_0(1500)}(q^2) & = (m_b +  m_s) (m_{B_s}  +  m_{f_0(1500)}) \, \frac{ \bar{f}_{f_0(1500)} }{m_{B_s}^2 f_{B_s}} \, e^{m_{B_s}^2/M^2} \int^{\tilde x_0}_{x_0} \, dx \, \exp \bigg[-\frac1{xM^2}
\nonumber
\\
&\times (m_b^2-\bar x q^2 + x\bar x m_{f_0(1500)}^2)\bigg] \bigg\{ -\frac{\phi_{2;f_0(1500)}(x)}{2x} + \frac{m_b m_{f_0(1500)}}{m_b^2 \! - \! q^2 \! + \! x^2m_{f_0(1500)}^2}
\nonumber
\\
&\times \bigg[ \frac{2xm_{f_0(1500)}^2}{m_b^2-q^2 + x^2m_{f_0(1500)}^2} - \frac{d}{dx} \bigg] \frac{\phi_{3;f_0(1500)}^\sigma(x)}{6} \bigg\}
\label{eq:LCSRTFFs3}
\end{align}
with
\begin{align}
x_0 &= \frac{ \sqrt{(q^2 - s_{B_s} + m_{f_0(1500)}^2)^2 + 4m_{f_0(1500)}^2 (m_b^2 - q^2)} + q^2 - s_{B_s} + m_{f_0(1500)}^2 }{2m_{f_0(1500)}^2},
\label{eq:x}
\\
\tilde x_0 &= \frac{\sqrt{(q^2 - \,\, {\cal T} \,\,  + m_{f_0(1500)}^2)^2 + 4m_{f_0(1500)}^2 (m_b^2 - q^2)} + q^2 - \,\, {\cal T} \,\, + m_{f_0(1500)}^2}{2m_{f_0(1500)}^2}.
\label{eq:xx}
\end{align}
Here $x_0$ and $\tilde x_0$ are the lower and upper integration boundaries determined by the on-shell condition of the $b$-quark within the light-cone sum rules for the $B_s\to f_0(1500)$ transition. The parameter $x_0$ is the critical momentum fraction obtained by adopting $m_{B_s}^2$, while $\tilde x_0$ follows from the same equation with $m_{B_s}^2$ replaced by the continuum threshold parameter ${\cal T}\equiv s_0^{B_s}$, and their explicit formulas are given in Eqs.~\eqref{eq:x} and Eqs.~\eqref{eq:xx}. In particular, the upper limit of the integral over $x$ in LCSR~\eqref{eq:LCSRTFFs1},~\eqref{eq:LCSRTFFs2} and~\eqref{eq:LCSRTFFs3} is $\tilde x_0$ instead of 1, because the lower limit of the integral variable $s$ in the dispersion relationship~\eqref{eq:DispersionRelationTFFs}, ${\cal T}$ is larger than but not equal to $m_b^2$.

The twist-2 LCDA of the $f_0(1500)$-resonance is the main source of nonperturbative uncertainty in our LCSR calculations. Since it is a universal, nonperturbative physical quantity, it is very appropriate to study it by combining nonperturbative quantum chromodynamics (QCD) with phenomenological models.
For the scalar resonance $f_0(1500)$, its twist-2 light-cone wave function $\Psi_{2;f_0(1500)}(x,\mathbf{k})$ can be expressed in terms of its rest-frame wave function $\Psi_{2;f_0(1500)}^{R}(x,\mathbf{k}_{\perp})$. To account for the spin projections ${\lambda_1\lambda_2}$ of the quark and antiquark, we introduce the spin projection coefficient $\chi_{2;f_0(1500)}^{\lambda_1\lambda_2}(x,\mathbf{k})$ as a bridge between the two frames, we establish a specific correspondence between the equal-time wave function in the rest frame and light-cone wave function in Refs.~\cite{Wu:2010zc,Wu:2011gf}, This ultimately leads to the final expression for the light-cone wave function of the $f_0(1500)$:
\begin{align}
\Psi_{2;f_0(1500)}(x,\mathbf{k}_{\perp}) = \sum_{\lambda_1\lambda_2} \chi_{2;f_0(1500)}^{\lambda_1\lambda_2}(x,\mathbf{k}_{\perp})\Psi_{2;f_0(1500)}^{R}(x,\mathbf{k}_{\perp}),
\end{align}
For the form of spin wave function$\chi_{f_0(1500)}^{\lambda_1\lambda_2}(x,\mathbf{k}_\perp)$, light meson wave function is usually transformed into the light-cone form to obtain complete spin wave function in the instantaneous SU(6) quark model~\cite{Huang:1994dy,Wu:2007rt,Ma:1993ht}. Starting from the instant-form, we take scalar state $ f_0(1500)$ with spin $S=1$, and orbital angular momentum $L=1$, and total angular momentum $J=0$. In the rest frame $(q_1 + q_2=0)$ the spin wave function in instant-form (T) can be obtained,
\begin{align}
\chi_{f_0(1500)}^{\rm T}=\frac{1}{\sqrt{2}}(\chi_1^\uparrow\chi_2^\downarrow-
\chi_2^\uparrow\chi_1^\downarrow),
\end{align}
where $\chi_{1/2}^{\uparrow/\downarrow}$ is the Pauli spinor of the triplet state, and the four-momenta of the two quarks are respectively: $q^{\mu}_1 =(q^0,\boldsymbol{q})$, $q^{\mu}_2 =(q^0,-\boldsymbol{q})$, $q^0=\sqrt{m^2 + \boldsymbol{q}^2}$. The instant-form $\vert J, s \rangle_T$ and the light-cone form $\vert J, \lambda \rangle_F$ are related by Wigner rotation, and for hadronic states with total $J=0$, this rotation reduces to the identity matrix,expressed as $|J,\lambda \rangle_F=\sum_s U_{s \lambda}^J |J,s \rangle_T$, for a quark with spin-1/2, the corresponding Melosh transformation is given as follows:
\begin{align}
\chi^\uparrow(T)=\omega[(q^ + + m)\chi^\uparrow(F)-q_R\chi^\downarrow(F)],
\nonumber \\
\chi^\downarrow(T)=\omega[(q^ + + m)\chi^\downarrow(F) + q_L\chi^\uparrow(F)],
\end{align}
where $w=[2q^ + (q^0 + m)]^{-1/2}, \quad q_{R/L}=q_1\pm iq_2, \quad q^ + =q^0 + q^3$. After substitution, the spin wave function of the $f_0(1500)$ state can be obtained,
\begin{align}
\chi_{f_0(1500)}(x,\mathbf{k}_\perp)=\sum_{\lambda_1,\lambda_2}C_0^F(x,\mathbf{k}_\perp,
\lambda_1,\lambda_2)\chi_{1}^{\lambda_1}(F)\chi_{2}^{\lambda_2}(F).
\end{align}
When expressed in terms of the instant-form momentum $q^\mu=\left(q^0,\boldsymbol{q}\right)$, the component coefficient $C_{0}^F(x,\mathbf{k}_\perp,\lambda_1,\lambda_2)$ can be calculated, and its specific form is as follows:
\begin{align}
C_0^F(x,\mathbf{k}_\perp,\uparrow,\downarrow)&= + \frac{m}{\sqrt{2(m^2 + \mathbf{k}_{\perp}^2)}}, \nonumber \\
C_0^F(x,\mathbf{k}_\perp,\downarrow,\uparrow)&=-\frac{m}{\sqrt{2\big(m^2 + \mathbf{k}_{\perp}^2)}}, \nonumber \\
C_0^F(x,\mathbf{k}_\perp,\uparrow,\uparrow)&=- \frac{(k_1-ik_2)}{\sqrt{2(m ^2 + \mathbf{k}_\perp^2)}}, \nonumber \\
C_0^F(x,\mathbf{k}_\perp,\downarrow,\downarrow)&=-\frac{(k_1 + ik_2)}{\sqrt{2(m^2 + \mathbf{k}_\perp^2)}},
\end{align}
these coefficients satisfy the following normalization relation: $\sum_{\lambda_1, \lambda_2} C_0^F(x, {\bf k}_\bot, \lambda_1, \lambda_2)^*$ $C_0^F(x, {\bf k}_\bot, \lambda_1, \lambda_2) = 1$. In addition, apart from the ordinary helicity component $(\lambda_1 + \lambda_2=0)$, there exist higher helicity components $(\lambda_1 + \lambda_2 = \pm 1)$, while the instant-form wave function contains only the ordinary helicity component. Then, the spin wave function can be defined as.
\begin{align}
\chi_{f_0(1500)}^{\lambda_1\lambda_2}(x,\mathbf{k}_\perp) = \frac{\hat{m}_q^2}{\sqrt{\mathbf{k}_\perp^2 + \hat{m}_q^2}}
\end{align}

Moreover, The BHL description establishes an equivalence between the spatial wave function $\Psi_{2;f_0(1500)}^R(x,\mathbf{k}_\perp)$ and the equal-time wave function, thereby allowing us to avoid the complex problem of solving an infinite set of coupled integral equations when determining the specific form of the wave function. This leads us to the following result:
\begin{align}
\quad \Psi_{2;f_0(1500)}^R(x,\mathbf{k}_{\perp}) = A_{2;f_0(1500)}\varphi_{f_0(1500)}(x) \exp \Big[ - \frac{\mathbf{k}_\perp^2 + \hat{m}_q^2}{8 \beta^2 {x}{\bar x }} \Big],
\end{align}
where the ${\bf k}_{\perp}$ denotes transverse momentum, $A_{2;f_0(1500)}$, $m_q^2$ denote normalization constant and light quark mass, respectively~\cite{Cao:1997hw,Huang:2004su}. According to Ref.~\cite{Zhong:2021epq}, the harmonic-oscillator exponential factor $\exp( - \mathbf{k}_\perp^2 - \hat{m}_q^2)/(8 \beta^2 {x}{\bar x })$ primarily controls broadening in the transverse momentum ${\bf k}_{\perp}$. For this, we can control the longitudinal distribution by introducing the function $\varphi_{2; f_0(1500)}(x)$, which allows adjustment of the $x$-direction shape (including endpoint behavior and width) of twist-2 LCDA for $f_0(1500)$ state. Usually, a form similar to the traditional Gegenbauer polynomial expansion is adopted. Based on this, we propose the following approach:
\begin{align}
\varphi_{2; f_0(1500)}(x)&=B_{2;f_0(1500)}C_1^{3/2}(2x-1).
\end{align}
 after integrating over the squared transverse $\mathbf{k}_{\perp}$, the final twist-2 LCDA of the $f_0(1500)$-resonance can be obtained, which can be expression as:
\begin{align}
\phi_{2; f_0(1500)}(x,\mu) &= \frac{A_{2;f_0(1500)} \hat{m}_q \beta_{2;f_0(1500)}}{4 \sqrt{2} \pi^{3/2}} \sqrt{x \bar x } \varphi_{2;f_0(1500)}(x)
\nonumber\\
&\times\left\{\mathrm{Erf}\left[\sqrt{\frac{\hat{m}_q^2 + \mu^2}{8\beta_{2;f_0(1500)}^2 x\bar x }} \right]-\mathrm{Erf}\left[\sqrt{\frac{\hat{m}_q^2}{8\beta_{2;f_0(1500)}^2 x\bar x }}\right]\right\}
\end{align}
where error function ${\rm Erf}(x) = 2\int^x_0 dt e^{-t^2}/\sqrt{\pi}$ is the error function,and we take $\hat{m}_q^2=370\ \rm{MeV}$~\cite{Hu:2024tmc}. The model parameters of the above amplitude schemes can be determined by the following criteria.

The average value of squared transverse momentum ${\bf k}_\perp^2$ in Ref.~\cite{Fu:2014uea}
\begin{align}
\langle\mathbf{k}_\perp^2\rangle = \frac{\int dxd^2\mathbf{k}_\perp|\mathbf{k}_\perp|^2|\psi_{2;f_0(1500)}(x,\mathbf{k}_\perp)|^2}{\int dxd^2\mathbf{k}_\perp|\psi_{2;f_0(1500)}(x,\mathbf{k}_\perp)|^2}, \hspace{-0.65cm}
\end{align}
which is consistent with the choice of Ref.~\cite{Wu:2010zc} for light diquark states or light mesons.

The Gegenbauer moments $a_n^{2;f_0(1500)}(\mu)$ can be derived from the LCDA within the following formula:
\begin{align}
a_n^{2;f_0(1500)}(\mu)=\frac{\int_0^1dx\phi_{2;f_0(1500)}(x,\mu)C_n^{3/2}(\xi)}{\int_0^1dx6x\bar x [C_n^{3/2}(\xi)]^2},
\label{gegenbauer}
\end{align}
where $\xi=(2x-1)$. Generally, the properties of the twist-2 LCDA are mainly determined by its first few terms.

We adopt the first-order Gegenbauer moment and average value of squared transverse momentum $\langle \mathbf{k}_\perp^2 \rangle$ to determine model parameters $A_{2;f_0(1500)}$ and $\beta_{2;f_0(1500)}$
\begin{align}
\label{yes}
\langle\xi^{n}_{2;f_0(1500)}\rangle|_{\mu}=\int_0^1 dx \, \xi^{n} \, \phi_{2;f_0(1500)}(x,\mu).
\end{align}

To better compare the LCHO model, we also adopt the truncated form of the $f_0(1500)$-resonance distribution amplitude and substitute it into our TFFs to compute the subsequent physical observables, where $a_0^{2;f_0(1500)}(\mu)$ is set to zero.~\cite{Cheng:2005nb},
\begin{align}
\phi_{2;f_0(1500)}^{\rm{TF}}(x,\mu)=6x\bar x \left[a_0^{2;f_0(1500)}(\mu) + \sum_{n=1}^{\mathcal{N}=3}a_n^{2;f_0(1500)}(\mu)
C_n^{3/2}(\xi)\right].
\end{align}

Finally, the $f_0(1500)$ state twist-3 LCDA can be generally expanded as a series of Gegenbauer polynomials. We take the first two odd-order moments of this truncated series, the specific form is as follows~\cite{Lu:2006fr, Han:2013zg, Braun:2003rp, Chernyak:1983ej}:
\begin{align}
\phi_{3;f_0(1500)}^{p}(x,\mu) &= 1 + \sum_{n=1}^{\mathcal{N}=2} a_{n,p}^{3;f_0(1500)}(\mu) C_{n}^{1/2}(\xi), \nonumber \\
\phi_{3;f_0(1500)}^{\sigma}(x,\mu) &= 6x\bar x \big[1 + \sum_{n=1}^{\mathcal{N}=2} a_{n,\sigma}^{3;f_0(1500)}(\mu) C_{n}^{3/2}(\xi) \big],
\end{align}
by choosing the above twist-3 distribution amplitudes as input parameters, we can obtain the complete TFFs.

\section{Numerical Analysis And Discussions}\label {sec:III}
For a systematic study of the rare quasi-four-body decay process $B_s \to f_0(1500)(\to \pi^ + \pi^-)\ell^ + \ell^-$, we adopt the latest data from the PDG~\cite{ParticleDataGroup:2026aaa} to determine the fundamental input parameters: the meson masses $m_{B_s}=5366.91\pm0.11~\rm{MeV}$, $m_{f_0(1500)}=1522\pm25~\rm{MeV}$, the quark masses, $m_b=4185.9\pm3.4~\rm{MeV}$, $m_s=93.5\pm0.5~\rm{MeV}$. The decay constant $f_{f_0(1500)}=490\pm50~\rm{MeV}$~\cite{Wang:2008da}, $f_{B_s}=225\pm5~\rm{MeV}$~\cite{FlavourLatticeAveragingGroupFLAG:2024oxs} at the initial energy scale $\mu_0 = 1\ \rm{GeV}$. For the $B_s \to f_0(1500)$ transition, the energy scale $\mu$ is chosen as the typical internal momentum of the $b$-quark inside the $B_s$ meson, to separate the short-distance perturbative contributions and long-distance non-perturbative effects encoded in the LCDAs. Consequently, the energy scale adopted in this work $\mu_k = ({m_{B_s}^2 - m_b^2})^{1/2} \simeq 3\ \text{GeV}$. As discussed in Section II regarding the twist-2 LCDA model parameters, we adopt $a_1^{2;f_0(1500)}(\mu_0) = -0.48$ and $\left\langle \mathbf{k}_\perp^2 \right\rangle^{1/2} = 0.375\ \mathrm{GeV}^2$.

Also, the model parameters of the LCDA scheme for the light scalar state $f_0(1500)$ are presented: $B_{2;f_0(1500)} = 0.55$, $A_{2;f_0(1500)} = 733.9$ and $\beta_{2;f_0(1500)} = -0.616$. The LCDA model contains three parameters $B_{2;f_0(1500)}$, $A_{2;f_0(1500)}$ and $\beta_{2;f_0(1500)}$, while only two independent moment constraints are available, namely the Gegenbauer moment $a_1^{2;f_0(1500)}(\mu_0)$ and the root-mean-square transverse momentum $\langle \boldsymbol{k}_\perp^2\rangle^{1/2}$. This leaves one degree of freedom in the model. Since $B_{2;f_0(1500)}$ is intrinsically coupled with $A_{2;f_0(1500)}$ and lacks direct experimental constraints, we take $B_{2;f_0(1500)}$ as this free parameter and fix its value in advance. With $B_{2;f_0(1500)}$ specified, the two moment equations form a closed system, from which the remaining two parameters $A_{2;f_0(1500)}$ and $\beta_{2;f_0(1500)}$ can be uniquely solved. We adopt $B_{2;f_0(1500)} = 0.55$ as our central working value to ensure consistency with our constructed LCDA. In order to evaluate the theoretical uncertainty induced by this free parameter, we let $B_{2;f_0(1500)}$ vary over the interval $[0.50, 0.60]$. This parameter window is chosen such that all values within this range yield physically acceptable LCDA profiles without unphysical behaviours, and it provides a conservative estimate of the uncertainty associated with our LCDA model.

In order to evaluate the theoretical uncertainty induced by this free parameter, we let $B_{2;f_0(1500)}$ vary over the interval $[0.40,\,0.70]$. This parameter window is chosen such that all values within this range yield physically acceptable LCDA profiles without un-physical behaviours, and it provides a conservative estimate of the uncertainty associated with our LCDA model.

\begin{figure}[t]
\begin{center}
\includegraphics[width=0.6\textwidth]{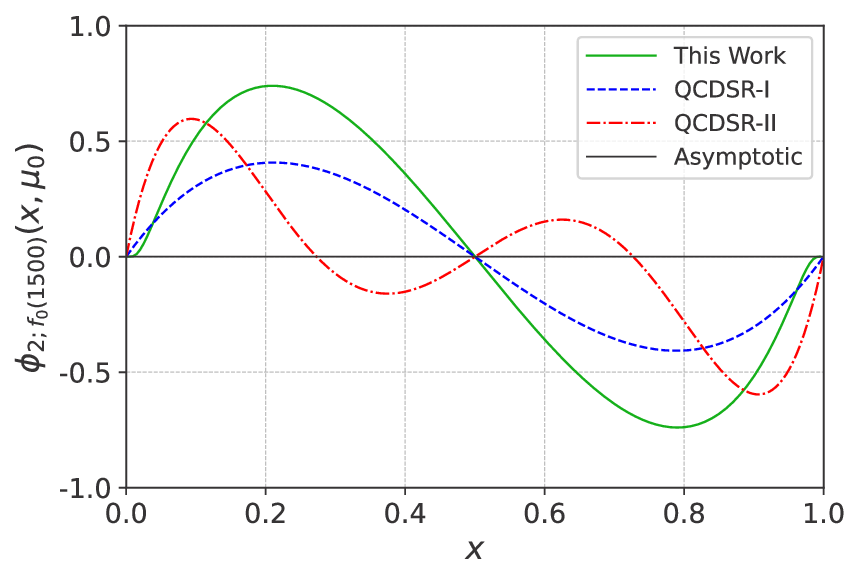}
\setlength{\tabcolsep}{6pt}
\end{center}
\caption{Behavior of twist-2 LCDA for $f_0(1500)$ at $\mu_0 = 1~\rm{GeV}$. As a comparison, the QCDSR-I and QCDSR-II (I and II in QCDSR~\cite{Sun:2010nv} correspond to the orders of Gegenbauer moments $N=1$ and $N=3$ respectively) are also presented.}
\label{Fig:DA}
\end{figure}

The corresponding behavior of the LCDA is shown in Fig.~\ref{Fig:DA}, where we can see that our prediction for the twist-2 LCDA exhibits an antisymmetric behavior, which is consistent with the results from QCD sum rules. As a comparison, the QCDSR-I and QCDSR-II (I and II in QCDSR~\cite{Sun:2010nv} correspond to the orders of Gegenbauer moments $N=1$ and $N=3$ respectively) are also presented. Our findings also agree in trend with those from QCDSR-I. However, there are some differences between them, which likely come from the different models used: we construct the twist-2 LCDA using the LCHO model, while QCDSR~\cite{Sun:2010nv} uses a Gegenbauer moment polynomial expansion. In addition, in our follow-up calculations, we found that the numerical results of QCDSR-I and QCDSR-II are almost identical. Therefore, in the following discussion, we will only use QCDSR-II for our analysis.

Using the definition Eq.~(\ref{yes}), we calculated the moments for $\langle\xi^{n}_{2;f_0(1500)}\rangle|_{\mu_0}$ with $(n=1,3,5)$ and the corresponding Gegenbauer moments. Since the zeroth-order Gegenbauer moment of the $f_0(1500)$-resonance is zero, and the higher-order even coefficients are strongly suppressed, all even-order coefficients vanish exactly under the approximation of equal constituent quark masses $(m_1=m_2)$. As a result, the twist-2 LCDA of $f_0(1500)$ is dominated by the odd-order Gegenbauer moments.
\begin{align}
&\langle\xi^{1}_{2;f_0(1500)}\rangle|_{\mu_0}=-0.281^{ + 0.092}_{-0.021},
~~~\hspace{0.2em}a_{1}^{2;f_0(1500)}(\mu_0)=-0.468^{ + 0.153}_{-0.035},
\nonumber \\
&\langle\xi^{3}_{2;f_0(1500)}\rangle|_{\mu_0}=-0.114^{ + 0.040}_{-0.033},
~~~\hspace{0.2em}a_{3}^{2;f_0(1500)}(\mu_0)=-0.034^{ + 0.003}_{-0.026},
\nonumber \\
&\langle\xi^{5}_{2;f_0(1500)}\rangle|_{\mu_0}=-0.059^{ + 0.022}_{-0.000},
~~~\hspace{0.2em}a_{5}^{2;f_0(1500)}(\mu_0)=-0.037^{ + 0.008}_{-0.008},
\end{align}
Meanwhile, we also present the $\langle\xi^{n}_{2;f_0(1500)}\rangle|_{\mu_k}$ moments and Gegenbauer moments at the corresponding energy scale $\mu_k =3~\rm{GeV}$,
\begin{align}
&\langle\xi^{1}_{2;f_0(1500)}\rangle|_{\mu_k}=-0.182^{ + 0.068}_{-0.003},
~~~\hspace{0.2em}a_{1}^{2;f_0(1500)}(\mu_k)=-0.300^{ + 0.110}_{-0.002},
\nonumber \\
&\langle\xi^{3}_{2;f_0(1500)}\rangle|_{\mu_k}=-0.072^{ + 0.028}_{-0.002},
~~~\hspace{0.2em}a_{3}^{2;f_0(1500)}(\mu_k)=0.032^{ + 0.003}_{-0.006},
\nonumber \\
&\langle\xi^{5}_{2;f_0(1500)}\rangle|_{\mu_k}=-0.037^{ + 0.015}_{-0.002},
~~~\hspace{0.2em}a_{5}^{2;f_0(1500)}(\mu_k)=0.016^{ + 0.011}_{-0.003},
\end{align}
In addition, the LCDA $\phi_{3;f_0(1500)}^{p,\sigma}(x,\mu)$ of $f_0(1500)$-resonance twist-3 distribution amplitude, we adopt the Gegenbauer moments at the energy scale $\mu_k $ as follows~\cite{Han:2013zg}:
\begin{align}
&a_{2,p}^{3;f_0(1500)}(\mu_k )=0.007_{-0.005}^{ + 0.019}, &a_{4,p}^{3;f_0(1500)}(\mu_k )&=0.227_{-0.165}^{ + 0.166},
\nonumber
\\
&a_{2,\sigma}^{3;f_0(1500)}(\mu_k )=-0.008\pm{0.007}, &a_{4,\sigma}^{3;f_0(1500)}(\mu_k )&=0.030 \pm{0.017}.
\end{align}

By substituting the above Gegenbauer moments, we can obtain the corresponding twist-3 distribution amplitudes. Using the twist-2 and twist-3 LCDA of the $f_0(1500)$-resonance we obtained, we can further calculate the TFFs for the $B_s \to f_0(1500)$ process. To determine the continuum threshold and Borel windows, we always take the two criteria:$(1)$ the contributions from the continuum and higher excited states are less than $30\%$; $(2)$ the dependence of form factors on the Borel parameter is weak. So we can determine the values of $s_0$ and $M^2$ to be $23.5 \pm0.5 {~\rm GeV}^2$ and $35 \pm1 {~\rm GeV}^2$ for $f_{\pm,\rm{T}}^{B_s f_0(1500)}$. With the above input parameters and the $f_0(1500)$-resonance DA parameters, we can obtain the values of the TFFs at zero momentum transfer, $i.e.$, $q^{2}=0\,\mathrm{GeV}^{2}$, and the moments with different input parameters, which are as follows.

\begin{align}
f_ + ^{B_s f_0(1500)}(0)& = + 0.390 \left(_{-0.023}^{ + 0.022}\right)_{s_{0}} \left(_{-0.006}^{ + 0.003}\right)_{M^2} \left( _{-0.003}^{ + 0.003}\right)_{m_{B_s}}  \left(_{-0.002}^{ + 0.002}\right)_{m_b} \left(_{-0.0006}^{ + 0.0009}\right)_{m_s}
\nonumber\\
& \quad + \left(_{-0.003}^{ + 0.003}\right)_{m_{f_0(1500)}} \left( _{-0.041}^{ + 0.040}\right)_{f_{f_0(1500)}} \left(_{-0.008}^{ + 0.009}\right)_{f_{B_s}}
\nonumber\\
&
=0.390_{-0.046}^{ + 0.047},
\label{Eq:fplusq2}
\\
f_-^{B_s f_0(1500)}(0)&=-0.460 \left(_{-0.023}^{ + 0.026}\right)_{s_{0}} \left(_{-0.003}^{ + 0.004}\right)_{M^2} \left( _{-0.004}^{ + 0.003}\right)_{m_{B_s}} \left(_{-0.002}^{ + 0.002}\right)_{m_b} \left(_{-0.0005}^{ + 0.0005}\right)_{m_s}
\nonumber\\
& \quad +\left(_{-0.010}^{ + 0.010}\right)_{m_{f_0(1500)}} \left( _{-0.047}^{ + 0.049}\right)_{f_{f_0(1500)}} \left(_{-0.010}^{ + 0.010}\right)_{f_{B_s}}
\nonumber\\
& = -0.460_{-0.055}^{ + 0.051},
\label{eq:f-}
\\
f_{\rm{T}}^{B_s f_0(1500)}(0)&= +0.568 \left(_{-0.031}^{ + 0.029}\right)_{s_{0}} \left(_{-0.004}^{ + 0.004}\right)_{M^2} \left( _{-0.005}^{ + 0.005}\right)_{m_{B_s}} \left(_{-0.003}^{ + 0.003}\right)_{m_b} \left(_{-0.0004}^{ + 0.0006}\right)_{m_s}
\nonumber\\
& \quad + \left(_{-0.005}^{ + 0.005}\right)_{m_{f_0(1500)}} \left( _{-0.058}^{ + 0.058}\right)_{f_{f_0(1500)}} \left(_{-0.012}^{ + 0.013}\right)_{f_{B_s}}
\nonumber\\
& =0.568_{-0.065}^{ + 0.069}
\label{eq:ft}
\end{align}

\begin{table*}[!h]
\renewcommand{\arraystretch}{1.5}
\begin{center}
\footnotesize
\caption{Numerical results for the $B_s\to f_0(1500)$ TFFs at large recoil region within errors. As a comparison, the theoretical predictions are also listed.}
\label{table:TFFsvalue}
\begin{tabular}{l l l l}
\hline
Method~~~~~~~~~~~~~~~ &$\ f^{B_s f_0(1500)}_ + (0)$~~~~~~~~~~~~~ &$\ f^{B_s f_0(1500)}_-(0)$~~~~~~~~~~~ &$\ f^{B_s f_0(1500)}_{\rm T}(0)$\\ \hline
This work        &$\ 0.390_{-0.046}^{ + 0.047}$        &$\ -0.460_{-0.055}^{ + 0.051}$        &$\ 0.568_{-0.065}^{ + 0.069}$\\
LCSR'10$a$~\cite{Han:2013zg}  &$\ 0.38_{-0.04}^{ + 0.04}$        &$\ -0.24_{-0.04}^{ + 0.04}$        &$\ 0.40_{-0.04}^{ + 0.04}$\\
LCSR'23~\cite{Han:2023pgf} &$\ 0.49_{-0.09}^{ + 0.09}$         &$\ -0.44_{-0.09}^{ + 0.09}$          &$\ 0.61_{-0.13}^{ + 0.13}$\\
LCSR'08~\cite{Wang:2008da}  &$\ 0.43$                &$\ -0.37$                &$\ 0.56$\\
LCSR'10$b$~\cite{Sun:2010nv}  &$\ 0.41$                &$\ -0.41$                &$\ 0.59$\\
pQCD'08~\cite{Li:2008tk} &$\ 0.60$         &$\ /$                &$\ 0.82$\\
\hline
\end{tabular}
\end{center}
\end{table*}

Meanwhile, we also include the predictions for TFFs $f_{\pm,\rm{T}}^{B_s f_0(1500)}(0)$ of the other approaches to compare with our results in Table~\ref{table:TFFsvalue}, such as the pQCD, and LCSR methods. As shown in Table~\ref{table:TFFsvalue}, the result for $f^{B_s f_0(1500)}_ + (0)$ is in good agreement with the central value from LCSR'10$a$, LCSR'08 and LCSR'10$b$ within a reasonable error range. However a comparison with pQCD'08 under different schemes reveals significant discrepancies in $f^{B_s f_0(1500)}_ + (0)$,  which can be primarily attributed to methodological differences.
Furthermore, our numerical results for $f^{B_s f_0(1500)}_-(0)$ and $f^{B_s f_0(1500)}_{\rm T}(0)$ agree well with LCSR'10 and LCSR'23, respectively. The physically allowable range for the TFFs is $0< q^2 < (m_{B_s} - m_{f_0})^2 \approx 14.78~\rm{GeV^2}$, but the LCSR approach for $B_s\to f_0(1500)$ TFFs are only reliable in low and intermediate region, $i.e.$, $q^2 \in [0,8]~\rm{GeV^2}$. In order to calculate the observables in phenomenology, such as branching fractions of the $B_s\to f_0(1500)\ell^ + \ell^-$ decays, we need extrapolate the TFFs in whole kinematical region $0< q^2 < (m_{B_s} - m_{f_0})^2$ via $z(q^2,t_0)$ converging the simplified series expansion (SSE)~\cite{Bharucha:2010im,Bharucha:2015bzk}. The TFFs take the following form

\begin{align}
f^{B_s f_0(1500)}_i(q^2) &= \frac{1}{1-q^2/m_{B_s}}\sum_{k=0,1,2}{\beta_kz^k(q^2,t_0)},
\label{eq:fi}
\end{align}
where $f^{B_s f_0(1500)}_i(q^2)$ with $i=( + ,-,{\rm T})$ represent the three TFFs. The $\beta_k$ are real coefficients and $z(q^2,t)$ is the function,
\begin{align}
z^k(q^2,t_0 ) =\frac{\sqrt{t_ + -q^2}-\sqrt{t_ + -t_0}}{\sqrt{t_ + -q^2} + \sqrt{t_ + -t_0}},
\label{eq:zk}
\end{align}
with $t_{\pm} = (m_{B_s} \pm m_{f_0(1500)})^2$ and $t_0=t_ + (1-\sqrt{1-t_-/t_ + })$. The most important is that we need to fit the appropriate value of three free parameters $\beta_{0,1,2}$ to make the quality of extrapolation $\Delta$ as small as possible. Then the quality of extrapolation $\Delta$ is defined as
\begin{align}
\Delta =\frac{\sum_t{|F_i(t)-F_i^{\rm fit}|}}{\sum_t{|F_i(t)|}}~\times 100.
\end{align}

\begin{figure*}[!h]
\begin{center}
\includegraphics[width=0.5\textwidth]{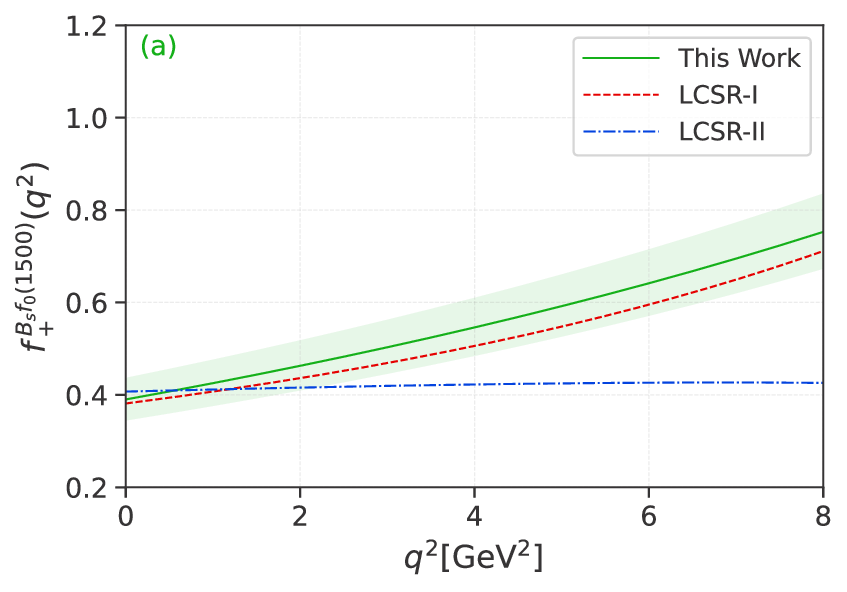}\includegraphics[width=0.5\textwidth]{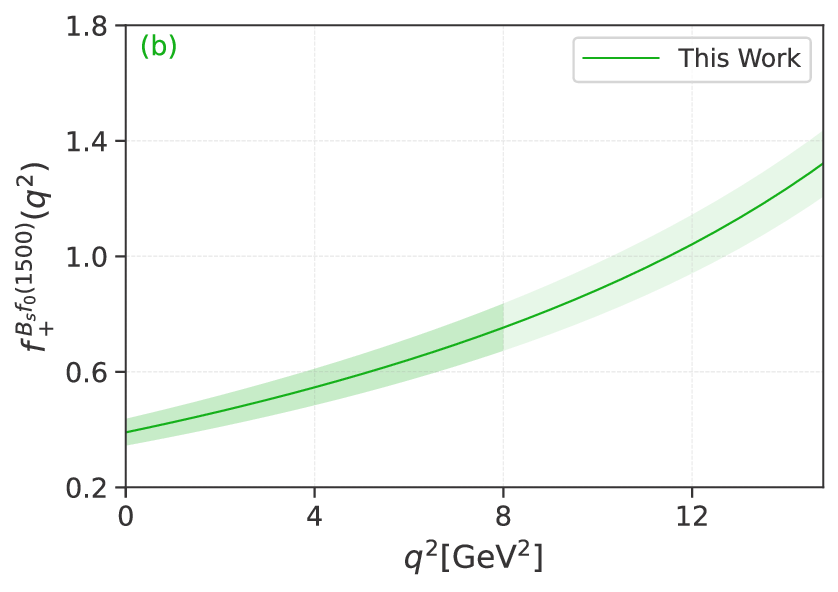}
\includegraphics[width=0.5\textwidth]{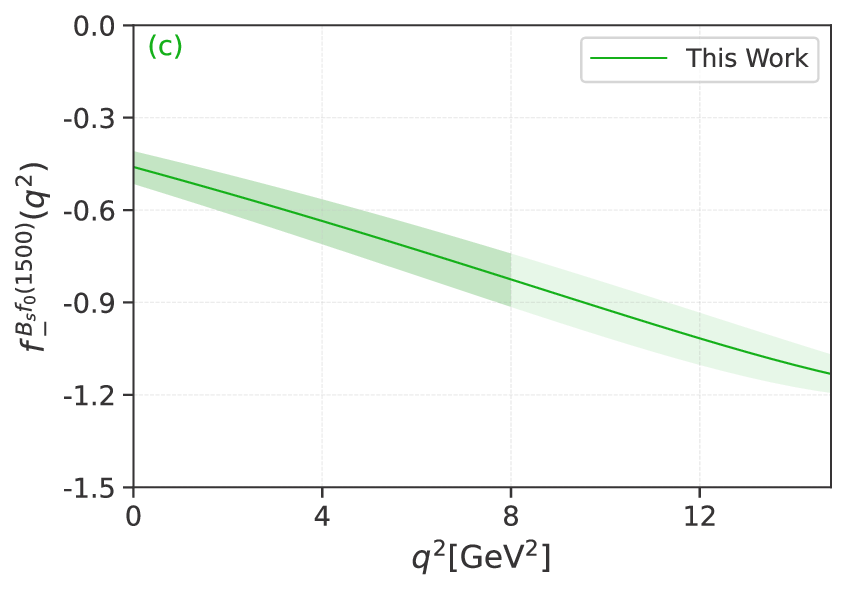}\includegraphics[width=0.5\textwidth]{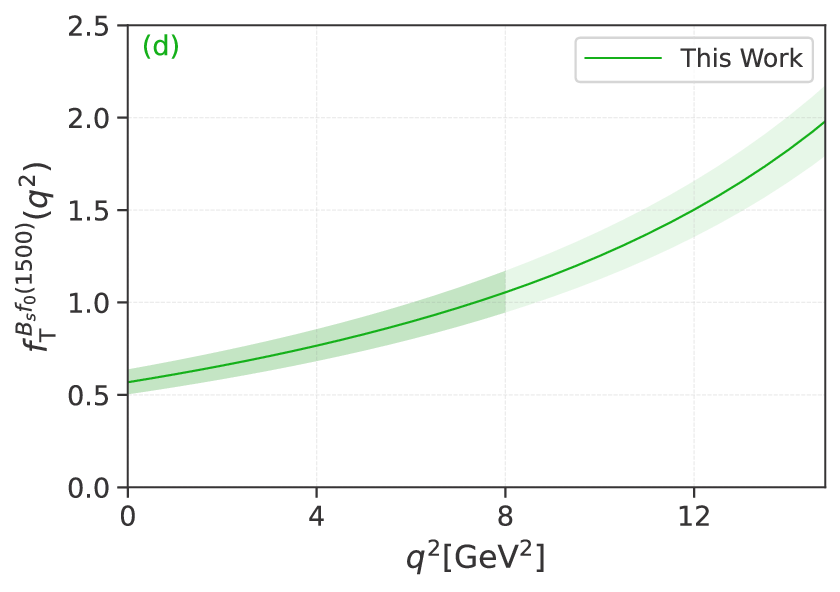}\\
\end{center}
\caption{Behaviors of the TFFs $f_ + ^{B_s f_0(1500)}$ in the low and intermediate $q^2$-region, planes (a) and the $f_{\pm,\rm{T}}^{B_s f_0(1500)}$ in whole $q^2$-region, plane (b), (c), (d). The solid line represents the central value and shaded bands corresponds to uncertainties. The lines in different colors represent theoretical predictions from LCSR-I~\cite{Han:2013zg}, LCSR-II~\cite{Sun:2010nv}. Their TFFs behaviors are used to compare with our results.}
\label{fig:TFFs}
\end{figure*}

\begin{table*}[!h]
\renewcommand{\arraystretch}{1.5}
\begin{center}
\footnotesize
\caption{The extrapolation parameters $\beta_{0,1,2}$ and quality of fit $\Delta$ for the $B_s\to f_0(1500)$ TFFs $f_{\pm,\rm{T}}^{{B_s f_0(1500)}} (q^2)$ based on the SSE approach.}
\label{table:parameters}
\begin{tabular}{lllll}
\hline
~~~~~~~~~~~~~~~~~~~~~~~&~~~~~~~~~~~~~&$ f_ + ^{{B_s f_0(1500)}}(q^2)$~~~~~~~~~~~&$ f_-^{{B_s f_0(1500)}}(q^2)$~~~~~~~~~~~~ &$ f_{\rm T}^{{B_s f_0(1500)}}(q^2)$\\ \hline
                      & $\beta_0$  &$ 0.437$               &$-0.409$                 &$ 0.637$\\
                      & $\beta_1$  &$-4.515$                &$4.835$                 &$-4.620$\\
\raisebox{2ex}[0pt]{upper limits}  & $\beta_2$  &$ -16.52$        &$33.421$                &$ 1.993$\\
                      & $\Delta$  &$ 0.934\%$              &$ -0.595\%$               &$ 0.116\%$\\ \hline
                      & $\beta_0$  &$ 0.390$               &$-0.460$                  &$ 0.568$\\
                      & $\beta_1$  &$-4.025$                 &$ 5.020 $                 &$-4.240$\\
\raisebox{2ex}[0pt]{central value}  & $\beta_2$  &$ -12.33$        &$ 42.703$                &$ 2.951$\\
                      & $\Delta$  &$ 0.886\%$              &$ -0.724\%$               &$ 0.112\%$\\ \hline
                      & $\beta_0$  &$ 0.344 $               &$-0.515$                  &$ 0.503$\\
                      & $\beta_1$  &$-3.601$                 &$ 5.712$                  &$-3.920$\\
\raisebox{2ex}[0pt]{lower limits}   & $\beta_2$  &$ -9.051$        &$ 53.149$                &$ 3.188$\\
                      & $\Delta$  &$ 0.873\%$              &$ -0.836\%$               &$ 0.134\%$\\ \hline
\end{tabular}
\end{center}
\end{table*}
After extrapolating the TFFs $f_i^{B_s f_0(1500)}(q^2)$ to the whole physical $q^2$ -region, we can obtain the coefficients $ \beta_{0,1,2} $ and $\Delta$, which are listed in Table~\ref{table:parameters}. We can observe that all the $\Delta$ values of $B_s \to f_0(1500)$ are no more than $1\%$, which indicate the effect of the extrapolation is perfect. The corresponding data in the Table~\ref{table:parameters} can help us to reproduce the results directly of TFFs by Eqs.~\eqref{Eq:fplusq2}, ~\eqref{eq:f-} and ~\eqref{eq:ft}. The behavior of the $ B_s \to f_0(1500) $ TFFs $f_{ + }^{B_s f_0(1500)}(q^2)$ in low and intermediate $q^2$ -region and the $f_{\pm,\rm{T}}^{B_s f_0(1500)}(q^2)$ in whole $q^2$ -region are shown in Fig.~\ref{fig:TFFs}, respectively. Since these literatures LCSR~\cite{Han:2013zg}, LCSR~\cite{Sun:2010nv}, depicted the behavior diagrams of the TFFs in the low and intermediate range of $0 < q^2 <8~{\rm GeV^2} $, for the sake of convenience, we hereby present their predictions within this region and subsequently compare them with our outcomes, Fig.~\ref{fig:TFFs} ({\color{blue}{a}}). The shadow band represents the error range, which is generated by the upper and lower limits of the input parameters. The solid line is our central result and other lines are the central predictions of other groups. Then, we can see that for $f^{B_s f_0(1500)}_ + (q^2)$, our result is in a good agreement with the results of LCSR-I~\cite{Han:2013zg} within reasonable error range.Our result differs from LCSR-II~\cite{Sun:2010nv} within the error range, which arises from the fact that we have included the contributions from higher-twist distribution amplitudes in our calculation, while they have not taken such higher-order contributions into account. In addition, we present the behavior of $f_{\pm,\rm{T}}^{B_s f_0(1500)}(q^2)$ with whole physical $q^2$-region in Fig.~\ref{fig:TFFs} ({\color{blue}{b}}), ({\color{blue}{c}}) and ({\color{blue}{d}}). We are more concerned about the extrapolation effect of $f^{B_s f_0(1500)}_ + (q^2)$ than $f^{B_s f_0(1500)}_{-,{\rm T}} (q^2)$, because in following calculations of observations, it makes a major contribution.

\begin{figure}[!h]
\begin{center}
\includegraphics[width=0.7\textwidth]{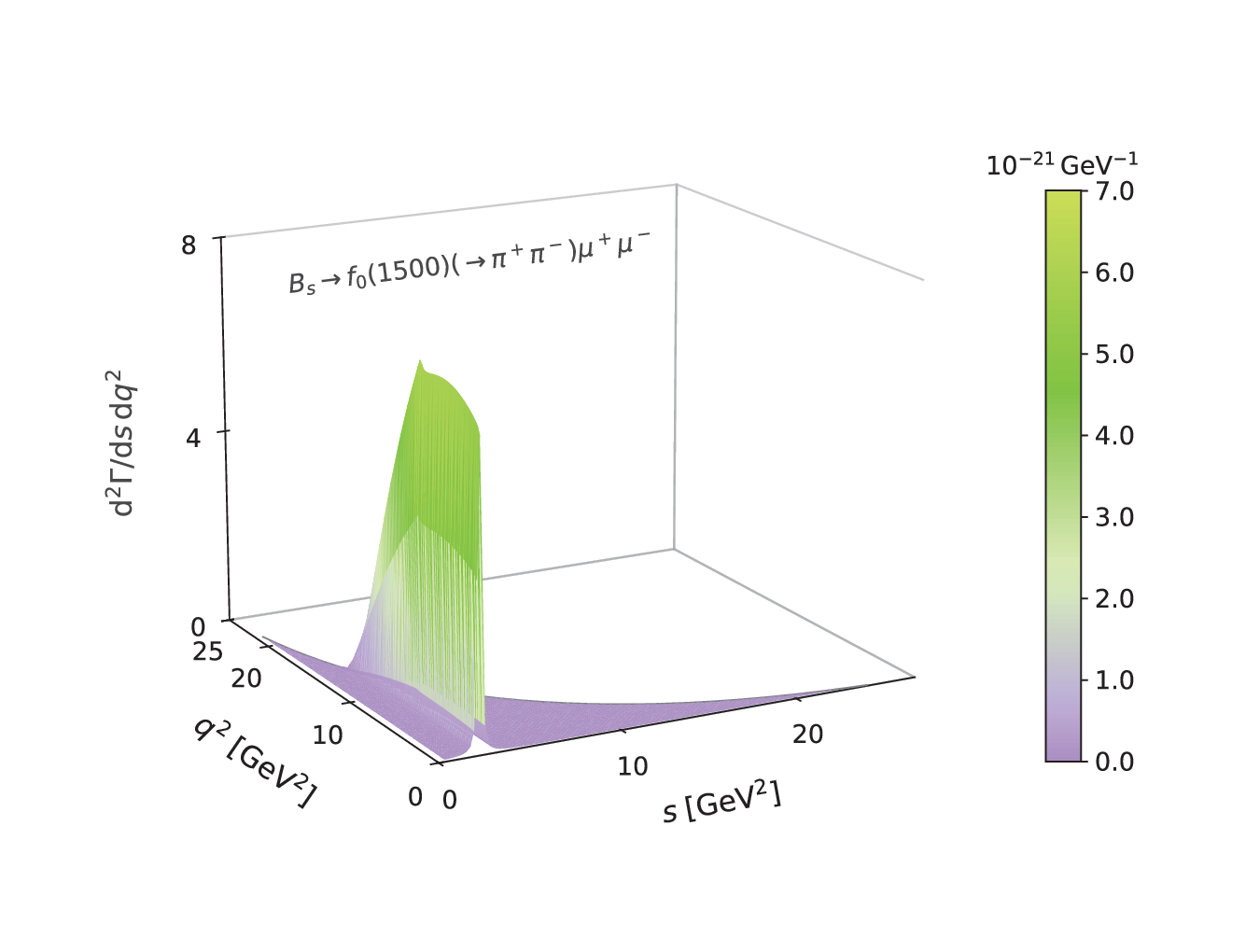}
\end{center}
\caption{Distribution of double-differential decay width $d^2\Gamma/ds dq^2$ for $B_s \to f_0(1500)(\to \pi^ + \pi^-) \mu^ + \mu^-$, as functions of $s$ and $q^2$. Both the height and color of the surface denote the value of $d^2\Gamma/ds dq^2$ in units of $10^{-21}\,\mathrm{GeV}^{-1}$.}
\label{Fig:dt}
\end{figure}

To show the distribution of this decay in phase space, we use Fig.~\ref{Fig:dt} to plot the 3D surface of the double-differential decay width $d^2\Gamma/dsdq^2$ for $B_s \to f_0(1500)(\to \pi^ + \pi^-)\mu^ + \mu^-$. This surface depends on $s$, the squared invariant mass of $\pi^ + \pi^-$, and $q^2$, the squared invariant mass of $\mu^ + \mu^-$. The height and color of the surface show the value of $d^2\Gamma/dsdq^2$, with units $10^{-21}\,\text{GeV}^{-1}$. A clear peak of the double-differential decay width can be seen in the $s$ range of the $f_0(1500)$-resonance. When $s$ moves away from this resonance area, the decay width drops quickly. The figure shows that the resonance increases the differential decay spectrum. It also shows that the finite width of $f_0(1500)$ changes the phase-space distribution.

Next step, combining the input parameters with Eqs.~(\ref{eq:DW1}) and (\ref{eq:DW2}), we can calculate the differential decay width of quasi-four-body rare $B_s \to f_0(1500)(\to\pi^ + \pi^-)\ell^ + \ell^-$ decay. In the calculation, respectively. Before performing the calculation, we need to define some parameters. Such as $m_\mu= 105.658~{\rm MeV}$, $m_e=0.511~{\rm MeV}$, $m_{\tau}= 1777.86~{\rm MeV}$, the meson masses $m_{\pi} = 0.13957~\rm{MeV}$, $m_{K} = 0.493677~\rm{MeV}$, the strong coupling: $g_{f_0(1500)\pi^ + \pi^-}=178.9~\rm{MeV}$ and $g_{f_0(1500)\ K^ + K^-}=100~\rm{MeV}$~\cite{Anisovich:2001ay}. fermi coupling constant $G_{\rm{F}}=1.166\times 10^{-5}~{\rm GeV^{-5}}$, CKM matrix elements $|V_{tb}|=0.9991$, $|V_{ts}|= 41.61\times 10^{-3}$, $\alpha_{\rm em}=1/137$. In addition, $C_7^{\rm eff}=-0.313, C_{10}=-4.669$~\cite{Grinstein:1988me}.
\begin{table}[!h]
\footnotesize
\begin{center}
\renewcommand{\arraystretch}{1.5}
\setlength{\tabcolsep}{4pt}
\caption{The branching fractions (in Unit: $10^{-7}$) of the quasi-four-body $B_s \to f_0(1500)(\to\pi^ + \pi^-)\ell^ + \ell^-$, $B_s \to f_0(1500)(\to \pi^ + \pi^-) \nu\bar{\nu}$ and the three-body $B_s \to f_0(1500)\ell^ + \ell^-)$, $B_s \to f_0(1500)\nu\bar{\nu}$ rare decays with $\ell=(e,\mu,\tau)$, respectively. For comparison, we also present the theoretical predictions of LCSR'08~\cite{Wang:2008da}, LCSR'14~\cite{Wang:2014upa}, pQCD'08~\cite{Li:2008tk} and LCSR'23~\cite{Han:2023pgf}.}
\label{table:BRs}
\begin{tabular}{l l l l l l}
\hline
Decay Modes &This Work &LCSR'08~\cite{Wang:2008da} &LCSR'14~\cite{Wang:2014upa} &pQCD'08~\cite{Li:2008tk} &LCSR'23~\cite{Han:2023pgf} \\
\hline
$B_s \to f_0(1500)(\to\pi^ + \pi^-)e^ + e^-$     &$5.047^{ + 1.117}_{-0.973}$ &$--$ &$--$   &$--$ &$--$\\
$B_s\to f_0(1500)(\to\pi^ + \pi^-)\mu^ + \mu^-$   &$5.022^{ + 1.113}_{-0.969}$  &$--$ &$--$   &$--$  &$--$\\
$B_s\to f_0(1500)(\to\pi^ + \pi^-)\tau^ + \tau^-$  &$0.096^{ + 0.033}_{-0.028}$ &$--$ &$--$   &$--$ &$--$\\
$B_s\to f_0(1500)(\to\pi^ + \pi^-)\nu\bar{\nu}$  &$28.25^{+6.75}_{-5.84}$ &$--$ &$--$   &$--$ &$--$\\
$B_s\to f_0(1500)e^ + e^-$            &$6.283^{ + 1.390}_{-1.211}$ &$5.3_{-1.8}^{ + 2.3}$ &$3.74(99)(2)$ &$10.0_{-3.8}^{ + 8.5}$  &$--$\\
$B_s\to f_0(1500)\mu^ + \mu^-$          &$6.251^{ + 1.386}_{-1.207}$ &$5.2_{-1.7}^{ + 2.3}$ &$3.72(99)(2)$ &$10.0_{-3.8}^{ + 8.5}$  &$--$\\
$B_s\to f_0(1500)\tau^ + \tau^-$         &$0.117^{ + 0.040}_{-0.034}$ &$0.12_{-0.05}^{ + 0.08}$ &$0.130(4)(0)$ &$0.13_{-0.06}^{ + 0.12}$  &$--$\\
$B_s\to f_0(1500)\nu\bar{\nu}$         &$35.17^{ + 8.41}_{-7.26}$ &$--$ &$--$ &$--$ &$26.7(101)$\\
\hline
\end{tabular}
\end{center}
\end{table}

Furthermore, utilizing the extrapolated TFFs within the SSE framework together with the lifetime of the $B_s$ meson, we evaluate the branching fractions for the quasi-four-body rare decay
$B_s \to f_0(1500)(\to \pi^ + \pi^-)\ell^ + \ell^-$ and $B_s \to f_0(1500)(\to \pi^ + \pi^-)\nu\bar{\nu}$ mediated by the $S$-wave $f_0(1500)$-resonance.
For comparison purposes, the branching ratios of the corresponding three-body rare decays
$B_s \to f_0(1500)\ell^ + \ell^-$ and $B_s \to f_0(1500)\nu\bar{\nu}$ are also computed under the narrow-width approximation.
All numerical results are summarized in Table~\ref{table:BRs}.

To the best of our knowledge, the quasi-four-body rare decay
$B_s \to f_0(1500)(\to \pi^ + \pi^-)\ell^ + \ell^-$ has not yet been systematically investigated in existing literature.
Motivated by this gap, we carry out a dedicated phenomenological analysis based on the branching ratios of the three-body rare-decay channel and the fundamental input parameters adopted in the previous studies.
The relativistic Flatt\'e parametrization is employed to perform our numerical evaluation, which enables a systematic comparison between different decay scenarios.
Our numerical results demonstrate that the branching fraction of the quasi-four-body decay, obtained by describing the subprocess $f_0(1500)\to\pi^ + \pi^-$ via the Flatt\'e formula, is considerably smaller than the prediction derived from the narrow-width approximation.
Relative to its three-body counterpart calculated under the narrow-width assumption, the quasi-four-body final-state configuration contains more decay products, which significantly restricts the available phase-space volume.
The decay channel considered in this work proceeds as a sequential process passing through the intermediate $f_0(1500)$-resonance, where the finite decay width associated with the transition $f_0(1500)\to\pi^ + \pi^-$ must be properly incorporated.

In addition, the Flatt\'e formula naturally accounts for final-state strong-interaction corrections and resonant effects within the $\pi^ + \pi^-$ system, rendering our theoretical description more consistent with the underlying physical reality.
Within the quoted uncertainty bands, our three-body results show good consistency with the predictions reported in LCSR'08~\cite{Wang:2008da}, LCSR'14~\cite{Wang:2014upa}, LCSR'23~\cite{Han:2023pgf}.
The deviations observed with respect to the and PQCD'08~\cite{Li:2008tk}, calculations can be mainly attributed to differences in the TFFs and the divergent calculation approaches among various models.
Constrained by lepton universality, the branching fractions of the electron and muon channels are almost equal. The very small difference between them comes from phase-space corrections caused by the tiny mass difference between electrons and muons. From the numerical values listed in Table~\ref{table:BRs}, one can see that the branching fraction of the quasi-four-body decay
$B_s \to f_0(1500)(\to \pi^ + \pi^-)e^ + e^-/ \mu^ + \mu^-$ is at the order of $5\times 10^{-7}$.
The $\tau^ + \tau^-$ decay channel is strongly suppressed by phase space, with its branching fraction only reaching the order of $10^{-9}$, which brings great difficulty to experimental detection. The branching fraction of the neutrino decay channel $B_s \to f_0(1500)(\to \pi^ + \pi^-)\nu\bar{\nu}$ is at the order of $10^{-6}$.
The numerical difference between three-body and quasi-four-body results clearly reflects the combined corrections from the finite width of the resonance, the final-state interaction of the $\pi\pi$ system, and phase-space reduction on the decay branching fractions. Such quasi-four-body rare decays are of great importance for testing the Standard Model and searching for new-physics signals beyond the Standard Model.

\section{Summary}\label{Sec:IV}
In this paper, we study the quasi-four-body rare decays $B_s \to f_0(1500)(\to \pi^ + \pi^-)\ell^ + \ell^-$ with $\ell=(e,\mu,\tau)$ and $B_s \to f_0(1500)(\to \pi^ + \pi^-)\nu\bar{\nu}$, which arise from the FCNC transition $b\to s$ and are known to be sensitive probes of the SM. Our aim is to obtain precise numerical results, so that the theoretical predictions of the Standard Model can be compared with future experimental observations. The $f_0(1500)$-resonance is treated as a quark-antiquark ground state. The TFF for $B_s\to f_0(1500)$ are calculated using the LCSR method. A key non-perturbative input is the twist-2 LCDA, which describes how momentum is shared inside the $f_0(1500)$-resonance. We model this LCDA by constructing one type of LCHO framework, the resulting behavior is shown in Fig.~\ref{Fig:DA}. In addition, we extract the Gegenbauer moments $a_{n;f_0(1500)}(\mu)$ at the scales $\mu_0$ and $\mu_k$, including the moments $\langle\xi^{n}_{2;f_0(1500)}\rangle |_{\mu}$. Based on the twist-2 LCDA, we then compute the form factor $f_ + ^{B_s f_0(1500)}$ at large recoil (q2=0) as well as its q2 dependence in the low and intermediate regions. Our results are consistent with the predictions from LCSR'08~\cite{Wang:2008da} and LCSR'10~\cite{Sun:2010nv} within the quoted uncertainties.

Then, we utilize TFFs to calculated the branching fractions of the quasi-four-body rare decays $B_s \to f_0(1500)(\to\pi^ + \pi^-)\ell^ + \ell^-$ and $B_s \to f_0(1500)(\to\pi^ + \pi^-)\nu\bar{\nu}$. For comparison, we also provide the results of the corresponding three-body decays under the narrow-width approximation.
To our knowledge, this quasi-four-body rare decay has not been systematically investigated before.
Our results show that the branching fraction of the four-body decay is considerably smaller the prediction obtained from the narrow-width approximation. Such difference is caused by the finite width of the $f_0(1500)$-resonance, final-state strong-interaction corrections and the reduced phase-space volume for quasi-four-body final states.
Our three-body results are in good agreement with most existing theoretical predictions LCSR'08~\cite{Wang:2008da}, LCSR'14~\cite{Wang:2014upa} and LCSR'23~\cite{Han:2023pgf}. The deviations among different studies mainly come from the differences of TFFs models and calculation schemes pQCD'08~\cite{Li:2008tk}. Constrained by lepton universality, the branching fractions of the electron and muon channels are almost equal. The $\tau^ + \tau^-$ channel is strongly suppressed by phase space, and the neutrino channel has the largest branching fraction among these decay modes.
Fig.~\ref{Fig:dt} shows the distribution of the double-differential decay width $d^2\Gamma/dsdq^2$ for $B_s \to f_0(1500)(\to \pi^ + \pi^-)\mu^ + \mu^-$. A prominent peak emerges in the resonance region of $f_0(1500)$, and the decay width falls rapidly away from resonance. It demonstrates the resonance enhancement and the modulation of phase-space distribution caused by the finite width of $f_0(1500)$.
The numerical difference between three-body and quasi-four-body results reflects the combined effects of resonance width, final-state interaction and phase-space reduction. These quasi-four-body rare decays are important for testing the Standard Model and searching for new-physics signals beyond the SM.

Overall, within the framework of the $q\bar{q}$ state assumption, our predictions for the physical observables
of the quasi-four-body rare decay $B_s \to f_0(1500)(\to\pi^ + \pi^-)\ell^ + \ell^-$ are reasonable and self-consistent. It should be noted,
however, that there are neither experimental measurements nor theoretical calculations available for this
quasi-four-body decay at present, so our results cannot be compared with any existing studies.
The internal structure of the scalar resonances family is still a controversial topic. The light scalar state
$f_0(1500)$ is particularly interesting in this respect. We hope that our predictions can serve as a useful
reference for future experimental searches of the quasi-four-body rare decay $B_s \to f_0(1500)(\to\pi^ + \pi^-)\ell^ + \ell^-$, and provide
some help for a deeper understanding of the internal structure of the light scalar state $f_0(1500)$.

\section{Achowledgements}
This work was supported in part by the National Natural Science Foundation of China under Grant No.12665017, No.12265010, the Project of Guizhou Provincial Department of Science and Technology under Grants No.MS[2025]219 and No.CXTD[2025]030.

\end{document}